\documentclass[fleqn,usenatbib]{mnras}
\usepackage{array} 
\usepackage{comment}
\usepackage{newtxtext,newtxmath}
\usepackage[T1]{fontenc}
\usepackage{booktabs}
\usepackage{ulem}

\DeclareRobustCommand{\VAN}[3]{#2}
\let\VANthebibliography\thebibliography
\def\thebibliography{\DeclareRobustCommand{\VAN}[3]{##3}\VANthebibliography}

\usepackage{graphicx}
\usepackage{amsmath}
\usepackage{makecell}

\title[IMBH growth in simulated dwarfs]
{Small hosts, big appetites: unveiling rapid and early low-mass black hole growth in cosmological zoom-in simulations of dwarf galaxies}

\author[G. Ortame et al.]{Giulia Ortame,$^{1,2}$\thanks{E-mail: go288@cam.ac.uk (GO)}
Martin A. Bourne,$^{2,3}$
Sophie Koudmani,$^{2,3}$
Debora Sijacki,$^{1,2}$ 
Francesco D'Eugenio$^{2,4}$\newauthor
and Roberto Maiolino$^{2,4}$ 
\\
$^{1}$Institute of Astronomy, University of Cambridge, Madingley Road, Cambridge, CB3 0HA, UK\\
$^{2}$Kavli Institute for Cosmology, Cambridge, University of Cambridge, Madingley Road, Cambridge, CB3 0HA, UK\\
$^{3}$Centre for Astrophysics Research, Department of Physics, Astronomy and Mathematics, University of Hertfordshire, College Lane, Hatfield AL10 9AB, UK\\
$^{4}$Cavendish Laboratory, University of Cambridge, 19 JJ Thomson Avenue, Cambridge CB3 0HE, UK
}

\date{MNRAS, accepted}

\pubyear{\the\year{}}

\begin{document}
\label{firstpage}
\pagerange{\pageref{firstpage}--\pageref{lastpage}}
\maketitle

\begin{abstract}
    Dwarf galaxies are ideal laboratories to probe the interplay between galaxy formation and the growth of black holes (BHs) in the early Universe. Mounting observational evidence reveals the presence of BHs in low-mass galaxies across cosmic time, with \textit{JWST} uncovering a likely population of \textit{overmassive} BHs at $2 \lesssim z \lesssim 11$. Simulations struggle to reproduce this high-redshift regime, motivating revisions to models of BH accretion and feedback from active galactic nuclei (AGN). To address this, we present high-resolution cosmological zoom-in simulations of a dwarf galaxy based on \textsc{fable} physics, introducing novel sink-based BH accretion models and relaxing the fiducial assumption of strong supernova feedback. BHs accrete more efficiently in the sink-based runs compared to the `traditional' Bondi-based counterparts, with AGN feedback leading to early, rapid quenching maintained by fast, hot and metal-enriched outflows. These outflows pollute the outer circumgalactic medium, yielding flat metallicity gradients down to $z=0$. We further assess the performance of two widely used virial estimators and find significant departures from the true dynamical mass, especially during the high-redshift dwarf assembly. Since our galaxy is dark-matter-dominated at all times and radii, BH growth, tied to the baryon cycle, shows no clear correlation with global dynamical properties. Efficient AGN feedback is produced by overmassive BHs relative to extrapolated local $M_\bullet - M_\star$ relations, raising the possibility that dormant, overmassive BHs in local quenched dwarfs and those probed by \textit{JWST} may reflect a common mode of early and rapid BH growth in low-mass galaxies. 
\end{abstract}




\section{Introduction}
\label{sec: introduction}

Massive black holes (BHs) are ubiquitous in our Universe across a broad range of masses and redshifts, with supermassive BHs ($M_\bullet \gtrsim 10^6 \; \mathrm{M}_\odot$) inhabiting the centres of virtually all local massive galaxies \citep[e.g. see][for reviews]{kormendy2013coevolution, greene2020intermediate}. As massive BHs are formed and reside in galactic centres, the physical processes governing their seeding and growth across cosmic time are deeply intertwined with those regulating galaxy formation and evolution in a cosmological context. Empirically, this co-evolutionary nature is manifested in well-established local scaling relations between BH masses and key host galaxy properties, including stellar bulge mass ($M_\bullet - M_\star$), stellar velocity dispersion ($M_\bullet - \sigma_\star$), dynamical mass ($M_\bullet - M_\mathrm{dyn}$) and luminosity ($M_\bullet - L$) \citep[e.g.][]{magorrian1998demography, ferrarese2000fundamental, gebhardt2000relationship, tremaine2002slope, kormendy2013coevolution, mcconnell2013revisiting, reines2015relations, shankar2019black, greene2020intermediate}. Theoretically, a coherent picture of the BH--host galaxy interplay, capable of explaining these empirical scaling relations, must be grounded in the standard $\Lambda-$Cold Dark Matter ($\Lambda$CDM) structure formation paradigm. In this framework, dark matter (DM) haloes originate from primordial density perturbations and then assemble hierarchically, forming the gravitational potential wells where primordial gas cools and galaxies are born \citep[e.g.][]{white1978core, blumenthal1984formation}. Within this hierarchical paradigm, dwarf galaxies ($M_\star \lesssim 3 \times 10^9 \, \mathrm{M}_\odot$) can be regarded as the building blocks of structure formation. As such, they represent ideal testbeds to probe the formation and co-evolution of BHs and their hosts. 

Dwarf galaxies in the local Universe may further provide a unique laboratory to shed light on the largely unconstrained pathways through which light and intermediate-mass BHs (IMBHs) with $M_\bullet \sim 10^2-10^6$ M$_\odot$ are seeded in the early Universe ($z>10$) and then evolve across cosmic time \citep[see][for reviews]{greene2020intermediate, inayoshi2020assembly}. Specifically, dwarf galaxies that evolved mostly at high redshift and in isolation may harbour central BHs, which may inform us about the properties and the seeding nature of BHs in the very early Universe \citep[e.g.][]{van2010massive, mezcua2017observational, pacucci2017feedback,  ricarte2018observational}. 

In these regards, the advent of the James Webb Space Telescope (\textit{JWST}) has opened an unprecedented window onto low-mass galaxies and their central BHs in the high-redshift Universe \citep[e.g.][]{ kocevski2023hidden, harikane2023jwst, ubler2023ga, maiolino2024small, maiolino2024jades, juodvzbalis2024dormant, juodvzbalis2025direct, juodvzbalis2026jades, matthee2024little,  ji2025blackthunder, d2025blackthunder, geris2026jades}, directly probing systems that may host the first BH seeds. Early \textit{JWST} results indicate that the $M_\bullet-M_\star$ relation may deviate significantly ($> 3\sigma$) from the local relation, revealing the presence of a likely population of \textit{overmassive} BHs, i.e. whose measured BH masses are a factor of $\sim 10-100$ larger relative to their local counterparts at comparable host stellar masses \citep[][]{pacucci2023jwst}. However, these observations may be influenced by selection effects, as high-redshift samples of active galactic nuclei (AGN) are typically biased towards the most luminous AGN relative to their host galaxies \citep[e.g.][]{li2025tip, geris2026jades}. Conversely, the $M_\bullet - \sigma_\star$ and $M_\bullet - M_\mathrm{dyn}$ relations seem to be more redshift invariant \citep[e.g.][]{maiolino2024jades}. 

Furthermore, due to their shallow gravitational potential wells, dwarfs are particularly susceptible to any perturbations, which can expel or heat the cold gas needed for forming stars. These include environmental processes (e.g. gravitational interactions, ram pressure stripping), stellar feedback \citep[e.g. supernova (SN) explosions, radiation, and winds from young stars;][]{hopkins2014galaxies, kimm2015towards, emerick2018stellar, smith2019cosmological}, and AGN feedback \citep[e.g. energetic winds, jets, and radiation from BH accretion;][]{sijacki2007unified, dubois2012self, bourne2017agn, costa2020powering}. Additionally, the metagalactic ultraviolet background can heat the primordial gas, suppressing star formation in low-mass dwarfs \citep[e.g.][]{efstathiou1992suppressing, benitez2015imprint}. Recent works indicate that cosmic rays (CRs) might also play a significant role, in some cases enhancing the efficiency of AGN feedback \citep[e.g.][]{su2021agn, su2024unravelling, wellons2023exploring, byrne2024effects, koudmani2025diverse}. 

Traditionally, reionisation and SN feedback have been invoked to regulate star formation at the low-mass end of the galaxy stellar mass function (GSMF), whereas AGN feedback is believed to be the dominant process at the high-mass end. While simulations that incorporate models for BH accretion and AGN feedback succeed in reproducing various key properties of the observed population of massive \textit{red and dead} galaxies \citep[e.g.][]{sijacki2015illustris, weinberger2016simulating,  weinberger2018supermassive, volonteri2016cosmic, henden2018fable}, models including only reionisation and SN feedback struggle to reproduce observations at the low-mass end, with some works indicating that additional physical processes (e.g. photoionisation, radiation pressure, CR physics and stellar winds) might be necessary for SN energy to effectively couple to the interstellar medium (ISM) and drive outflows \citep[e.g.][]{emerick2018stellar, smith2019cosmological, martin2023pandora, martin2026pandora, rey2025edge}. Moreover, accounting for realistic baryonic feedback processes in dwarfs could also significantly mitigate some of the small-scale challenges to the successful $\Lambda$CDM cosmological model \citep[see e.g.][for comprehensive reviews]{bullock2017small, sales2022baryonic}.

Beyond these theoretical considerations, mounting observational evidence for IMBHs and AGN activity in dwarfs calls this standard paradigm into question \citep[e.g.][]{silk2017feedback}. Signatures of gas accretion onto BHs in dwarfs have been detected across the entire electromagnetic spectrum, including X-ray \citep[e.g.][]{latimer2021agn, birchall2022incidence, sacchi2024x}, optical \citep[e.g.][]{molina2021sample, polimera2022resolve, Pucha2025}, infrared \citep[e.g.][]{marleau2017infrared, kaviraj2019agn, bichang2024properties} and radio wavelengths \citep[e.g.][]{mezcua2019radio, davis2022radio}. Furthermore, a population of overmassive BHs in low-mass galaxies out to $z \sim 3$ has been identified in the VIPERS survey \citep[e.g.][]{siudek2023environment, mezcua2023overmassive, mezcua2024overmassive}, providing an interesting link with the high-redshift overmassive BHs probed by \textit{JWST}. The environmental conditions needed to trigger AGN activity in dwarf galaxies are still poorly understood \citep[e.g.][]{kristensen2021merger, erostegui2025manga}, and whether the AGN is able to drive impactful outflows remains a hotly debated topic, but there is an increasing body of promising observational hints \citep[e.g.][]{penny2018sdss, manzano2019agn, mezcua2019radio, mezcua2020hidden, liu2020integral, zheng2023escaping, liu2024fast, wang2025rubies, morales2025manga}. 

These observational signatures have motivated the inclusion of BH accretion and AGN feedback modelling in theoretical studies of dwarf galaxies, albeit with mixed outcomes. While analytical models generally predict comparable AGN and SN feedback energetics in low-mass galaxies hosting IMBHs \citep[e.g.][]{silk2017feedback, dashyan2018agn}, some hydrodynamical simulations indicate that efficient SN feedback in galaxies with $M_\star \lesssim 10^9 \; \mathrm{M}_\odot$ inhibits the infall of dense, cold gas near the BH, suppressing its growth and AGN activity \citep[e.g.][]{dubois2015black, angles2017black, habouzit2017blossoms, trebitsch2018escape, hopkins2022black, byrne2023stellar, petersson2025noctua}. However, recent high-resolution simulations of isolated dwarfs with multi-phase ISM modelling show that nuclear star clusters (NSCs) may play a crucial role in funnelling gas towards the central regions, thereby promoting sustained accretion onto the BH \citep[e.g.][]{partmann2025importance, shin2025mandelzoom, shin2026mandelzoom2}. Furthermore, both observations and simulations indicate that IMBHs may wander in the shallow potentials of their hosts, displacing them from dense, gas-rich centres needed for efficient accretion \citep[e.g.][]{mezcua2020hidden, bellovary2021origins, sharma2022hidden, beckmann2023population}. Simulations also struggle to reproduce the observed AGN occupation fraction in low-mass galaxies, which is relatively well-constrained for X-ray bright sources \citep[e.g.][]{mezcua2018extended, birchall2020x}. Primarily, predictions are highly sensitive to the combined choice of the initial BH seed mass and the BH accretion prescription adopted: `heavy' seed simulations like \textsc{romulus} and IllustrisTNG tend to overproduce bright AGN, while `lighter' seed simulations such as \textsc{eagle}, \textsc{fable} and Illustris underproduce them \citep[e.g.][]{koudmani2021little, sharma2022hidden, haidar2022black}. Simplistic Bondi-based BH accretion models are partly responsible for this mismatch, as their quadratic dependence on the BH mass penalises the growth of low-mass BHs compared to gas-supply- and torque-limited schemes \citep[e.g.][]{angles2013black, angles2015torque, koudmani2022two, wellons2023exploring, gordon2024hungry, weinberger2025accretion}. Strong SN feedback parametrisations, used to match the low-mass end of the GSMF, further curtail BH growth. Nevertheless, some simulations, adopting different modelling choices, show that BHs in dwarfs can accrete efficiently and drive outflows that suppress star formation, provided that sufficient gas is present in the central regions of their hosts \citep[e.g.][]{barai2019intermediate, koudmani2019fast, koudmani2021little, koudmani2022two, sharma2020black, sharma2022hidden, sharma2023active, wellons2023exploring, arjona2024role}. 

In this work, we introduce a viable subgrid alternative to Bondi-based BH accretion models, namely a novel \textit{sink-particle} method \citep[see e.g.,][]{power2011accretion}, to model BH accretion in cosmological zoom-in simulations based on \textsc{fable} \citep{henden2018fable} physics, relaxing the fiducial assumption of strong SN feedback. Our primary aim is to perform computationally efficient, simple numerical experiments that allow us to access regions of parameter space in BH growth and AGN feedback efficiency that are challenging to probe within the constraints of conventional Bondi-based accretion models typically adopted in large-scale cosmological simulations, especially for low-mass galaxies in the early Universe. In this sense, the sink-based and Bondi-based models explored here should not be interpreted as competing physical descriptions, but rather as complementary limiting cases to help quantify the modelling uncertainty associated with unresolved BH accretion physics. This is especially relevant in the context of recent \textit{JWST} discoveries, which are beginning to probe BH growth in low-mass systems at high redshift, where theoretical predictions are expected to be particularly sensitive to subgrid modelling assumptions. 

The basic simulation setup, BH accretion and baryonic feedback models are described in Section~\ref{sec: methodology}. In Section~\ref{sec: cosmic evolution}, we analyse the evolution of BH accretion properties, star formation rates (SFRs), stellar and gas masses across cosmic time. In Section~\ref{sec: Mdyn estimators}, we examine the performance of two commonly used virial estimators, i.e. the \citet{walker2009universal} and the \citet{van2022mass} estimators, in recovering the true dynamical masses of our simulated dwarf galaxies across cosmic time. In Section~\ref{sec: scaling relations}, we compare our simulated dwarfs to key empirical scaling relations between BH and host galaxy properties, namely the $M_\bullet-M_\star$, $M_\bullet - \sigma_\star$ and $M_\bullet-M_\mathrm{dyn}$ relations. In Section~\ref{sec: outflows} we analyse the kinematic and metal-enrichment properties of AGN-driven outflows, and we briefly discuss the possibility of constraining the BH growth history and past AGN feedback activity through the metal enrichment of the outer circumgalactic medium (CGM). Finally, in Section~\ref{sec: discussion} we discuss our results in the context of current and future observational and theoretical studies, and we summarise our conclusions in Section~\ref{sec: conclusions}. 


\section{\textsc{Methodology}}
\label{sec: methodology} 


\subsection{Basic simulation properties}
\label{sec: sims properties}

The simulations analysed in this work are based on the \textsc{fable} suite
of cosmological simulations of galaxies, groups and clusters \citep{henden2018fable}. These simulations use the massively parallel gravity and magnetohydrodynamics code \textsc{arepo} \citep{springel2010arepo}, in which the equations of hydrodynamics are discretised and solved on a moving Voronoi-mesh using a quasi-Lagrangian finite volume approach. Gravitational interactions are treated using a tree-particle-mesh (TreePM) method, with stars, DM, and BHs modelled as collisionless particles. 

The initial conditions are generated with the \textsc{music} code \citep{hahn2011multi} at $z=127$, using the cosmological parameters from Planck Collaboration XIII \citep{ade2016planck}. The DM-only coarse simulation box has a size of 10 $\mathrm{cMpc} / h$, and the number of DM particles is 256$^3$, i.e. the mass resolution of the coarse simulation is $7.47\times 10^6$ M$_\odot$. At $z=0$, the target dwarf halo selected from this simulation has a virial mass $M_\mathrm{vir}= 1.04\times 10^{10} \; \mathrm{M_\odot}$ and a virial radius $R_\mathrm{vir}=62.0$ kpc\footnote{The viral radius and virial mass are hereby defined as $R_{200}$ and $M_{200}$, i.e. the radius and mass of a sphere centred on the particle with the minimum gravitational potential energy and whose mean density is 200 times the critical density of the Universe, at the time the halo is considered.}. 

For our suite of zoom-in simulations, the resolution is increased by a factor of 16$^3$, and the high-resolution simulation volume is restricted to a sphere of radius 736 kpc at $z=0$ enclosing the dwarf halo, yielding a high-resolution DM particle mass of $m_\mathrm{DM}=1536$  M$_\odot$ and a high-resolution target gas mass of $\overline{m}_\mathrm{gas}=287$ M$_\odot$. The comoving gravitational softening lengths of collisionless particles are initially set to $\epsilon_\mathrm{soft}=0.129\; \mathrm{ckpc}$ and  kept constant in physical units from $z=2$ onwards. Gravitational softening lengths for gas cells  instead evolve adaptively in proportion to the cell size, down to a minimum comoving softening length of $\epsilon_\mathrm{soft,min}=0.0431 \; \mathrm{ckpc}$. 

Note that this dwarf is the same system analysed in \citet{smith2019cosmological} (\textit{Dwarf 1}) and in \citet{koudmani2022two}. In this work, the initial conditions are left unchanged, whereas the BH accretion and baryonic feedback prescriptions are modified by introducing novel sink-based models for BH accretion.


\subsection{Star formation, ISM and stellar feedback}
\label{sec: methods SF and ISM}

Star formation and chemistry are modelled as in \textsc{fable}, which is largely based on the original Illustris implementation. In this section, we only report the main aspects of this modelling and refer the reader to \citet{vogelsberger2013model} and \citet{henden2018fable} for further details.

The ISM is unresolved and treated in a subgrid fashion via an effective equation of state (eEOS), following the approach described in \citet{springel2003cosmological}. Within this framework, star formation occurs stochastically above a density threshold, i.e. $\rho_\mathrm{sfr} = 0.13 \; \rm{cm}^{-3}$, and with a characteristic star formation timescale, i.e. $t_\mathrm{sfr} = 2.2 \; \rm{Gyr}$. The rate of energy input from SN explosions is computed based on a Chabrier initial mass function (IMF) \citep{chabrier2003galactic}. In addition to thermally regulating the ISM, a fraction of energy from SN explosions drives kinetic winds, where in our study the wind energy factor $\epsilon_\mathrm{W,SN}$ is set to $0.5$. Note that \citet{koudmani2022two} benchmarked this value against the fiducial \textsc{fable} ($\epsilon_\mathrm{W,SN}=1.5$) and Illustris ($\epsilon_\mathrm{W,SN}=1.09$) models. This choice allows us to explore whether a combination of mild SN feedback and novel BH accretion models can be a viable alternative. Furthermore, while in Illustris the energy carried by wind particles is purely kinetic, in \textsc{fable} one third of the wind energy is thermal, such that the ejected gas is characterised by longer cooling times and can remain outside of the star-forming ISM for longer.


\subsection{Black hole accretion and AGN feedback}
\label{sec: methods BH accretion and AGN feedback}


\subsubsection{Black hole seeding} 
\label{sec: methods BH seeding}

BHs are modelled as collisionless particles that are allowed to grow only by accretion of ambient gas. A single BH particle is seeded on-the-fly at a target redshift $z_\mathrm{seed}$ by converting the gas cell with the highest density in the most massive halo into a BH particle. Therefore, we do not model BH-BH mergers in this work. 

In order to explore the effects of different seeding redshifts and BH seed masses on the co-evolution of the BH and its host galaxy, as well as to allow for direct comparison with both local and high-redshift observations and empirical scaling relations, we consider two seeding configurations: a \textit{Low-z} seeding setup ($z_\mathrm{seed} \sim 4$ and $M_\mathrm{seed} \sim 10^4 \, \mathrm{M}_\odot$) and a \textit{High-z} seeding setup ($z_\mathrm{seed} \sim 15$ and $M_\mathrm{seed} \sim 10^3 \, \mathrm{M}_\odot$). Each BH accretion model is then run for both configurations, as described in Section~\ref{sec: sims overview} and summarised in Table~\ref{tab: runs}. 

Although the pathways through which BHs are seeded in the early Universe are still uncertain \citep[see][for a review]{inayoshi2020assembly}, the BH seed masses and seeding redshifts chosen in this work likely correspond either to the \textit{light-seed} scenario, i.e. the BH formed as the remnant of a Population III star, or to the i\textit{ntermediate-mass-seed} scenario, i.e. the BH formed as a result of gravitational runaway interactions in dense star clusters. Note that we do not consider BHs with higher seed masses that would be formed via direct collapse, as this seeding mechanism is most likely only relevant for higher galaxy masses.


\subsubsection{Black hole accretion: Bondi-based model}
\label{sec: methods Bondi-based accretion}

In Bondi-based BH accretion models, the BH accretion rate is dictated by a (boosted) Bondi-Hoyle-Littleton prescription \citep{bondi1944mechanism}. The accretion rate is thus computed as
\begin{equation}
    \label{eq: Bondi accretion}
    \dot{M}_\mathrm{acc} = \alpha \dot{M}_\mathrm{Bondi} = \alpha \frac{4\pi \mathrm{G}^2 M^2_\bullet \rho}{c^3_\mathrm{s}} \, ,
\end{equation}
where G is the gravitational constant, $M_\bullet$ is the mass of the BH, and $\rho$ and $c_\mathrm{s}$ are the local density and sound speed of the surrounding medium, respectively. A fiducial boost factor $\alpha=100$ is introduced to compensate for the unresolved multi-phase structure of the ISM \citep[e.g.][]{springel2005modelling, booth2009cosmological}. The local properties of gas in the vicinity of the BH are estimated on-the-fly within the BH smoothing length, i.e. the region which contains a total gas mass equal to 32 times the target gas cell mass. Following the approach of \citet{koudmani2022two}, the BH smoothing length has a ceiling of three times the comoving gravitational softening length of the BH, i.e. $r_\mathrm{max}=3 \times \epsilon_\mathrm{soft,\bullet}$.

The most striking feature of the Bondi-based prescription is its quadratic dependence on the BH mass, which by definition suppresses the growth of low-mass BHs, making it particularly challenging to assess the extent to which AGN feedback can impact the evolution of its dwarf host within the constraints of this model. Therefore, in the next section we introduce a novel sink-particle prescription to relieve this numerical limitation.


\subsubsection{Black hole accretion: sink-based model}
\label{sec: methods sink-particle model}

In addition to the traditional Bondi-based model, we present a novel BH accretion scheme that does not numerically penalise the growth of BHs in low-mass galaxies. Due to current computational limitations, it is not possible to resolve the accretion disc around the BH in our cosmological zoom-in simulations. Therefore, our subgrid model describes BH accretion through an effective collisionless \textit{sink particle}, whose accretion rate is computed directly from the resolved gas distribution within a finite sink region surrounding the BH.
This idea follows the approach originally developed in \citet{power2011accretion}, although we do not explicitly track the BH and accretion disc as separate components. 
The mass accreted by the sink particle is computed on-the-fly as a Gaussian-weighted sum of the masses of all gas cells within a given sink radius $R_\mathrm{sink}$. Hence, the accretion rate is computed as
\begin{equation}
\label{eq: sp Mdot}
    \dot{M}_\mathrm{acc} = \frac{M_\mathrm{drain}}
{\mathrm{d}t} \,,
\end{equation}
where $\mathrm{d}t$ is an assumed accretion timescale (see Section~\ref{sec: sims overview}), and 
\begin{equation}
\label{eq: Mdrain}
    M_\mathrm{drain} = \sum_{r<R_\mathrm{sink}}m_\mathrm{cell} \omega(r)\, .
\end{equation}
Here, $\omega(r)$ is a Gaussian function capped at 0.9 and centred on the BH, i.e. 
\begin{equation}
    \label{eq: Gaussian kernel}
        \omega(r) = \min \Biggl[ 0.9, \; \exp\Bigg(-4 \frac{r^2}{R_\mathrm{sink}^2} \Bigg)\Biggr] \; ,
\end{equation}
where $r$ is the distance of a given gas cell from the BH particle. Following the approach described in \citet{power2011accretion}, $R_\mathrm{sink}$ is treated as a free parameter in the simulations. 

In this work, we set the sink radius to $R_\mathrm{sink} = 0.387 \; \rm{ckpc}$ at all times, which is equal to three times the initial comoving gravitational softening length of the BH. This choice ensures that the region from which the BH accretes is always reasonably well resolved in the simulations. 


\subsubsection{AGN feedback}
\label{sec: methods AGN feedback}

A fraction ($1-\epsilon_\mathrm{r}$) of the accreted mass is added to the BH mass, such that $\dot{M}_\bullet = (1-\epsilon_\mathrm{r})\dot{M}_\mathrm{acc}$, where $\epsilon_\mathrm{r}=0.1$ is the radiative efficiency, i.e. the mean value for the radiatively efficient \citet{shakura1973black} accretion onto a BH. The remaining rest mass energy is returned as AGN luminosity, i.e. $L_\mathrm{bol}=\epsilon_\mathrm{r}\dot{M}_\mathrm{acc}\mathrm{c}^2$. As in \citet{koudmani2022two}, it is assumed that a fraction $\epsilon_\mathrm{f}=0.1$ of the AGN luminosity couples to the ISM and is injected as purely thermal AGN feedback, instead of distinguishing between the two accretion-dependent modes of AGN feedback as in the original \textsc{fable} and Illustris models. This feedback is injected isotropically within the BH smoothing length, limited to $r_\mathrm{max}$ for the Bondi-based runs and to $R_\mathrm{sink}$ for the sink-based runs, and weighted by the gas cell masses. For the majority of the simulation runs explored in this work, the feedback energy is continuously injected throughout the simulation. Conversely, in the \textit{Sink+Duty} model (see Section~\ref{sec: sims overview}), the feedback is injected following a duty cycle of $\Delta t_\mathrm{AGN} = 25 \, \mathrm{Myr}$, like in the original \textsc{fable} implementation. This means that the feedback energy is accumulated over $\Delta t_\mathrm{AGN}$ and then released in individual bursts, resulting in more efficient AGN feedback owing to the higher relative energy injected into the surrounding gas cells. This implementation is based on the approach of \citet{booth2009cosmological}, with the aim of avoid artificial overcooling \citep[e.g.,][]{schaye2015eagle, bourne2015resolution}, which can occur when cells do not reach sufficiently high temperatures required to avoid short cooling times.


\subsection{Overview of the simulation runs}
\label{sec: sims overview}

In this work, we examine one Bondi-based model and three sink-based models, in order to explore the sensitivity and robustness of our novel BH accretion prescription to changes in its parameters, as well as to compare its performance against common Bondi-based approaches. For each of these models, we run one \textit{Low-z} and one \textit{High-z} seeding simulation (see Section~\ref{sec: methods BH seeding}), which we label with the suffixes \textit{+Low-z} and \textit{+High-z}, respectively. Table~\ref{tab: runs} provides an overview of the characteristic parameter choices for all simulation runs.

\begin{table*}
    \centering
    \renewcommand{\arraystretch}{1.5}
    \begin{tabular}{l | c c c c c}
\toprule
  \makecell{\textbf{Simulation run}} & 
  \makecell{\textbf{BH accretion mode}} & 
  \makecell{\textbf{Sink radius} \\ $R_\mathrm{sink}$ [ckpc]} & 
  \makecell{\textbf{BH seeding} \textbf{redshift} \\ $z_\mathrm{seed}$} & 
  \makecell{\textbf{Duty cycle} \\ $\Delta_\mathrm{AGN}$ [Myr]} & 
  \makecell{\textbf{BH accretion} \textbf{timescale} \\ (Eq. \ref{eq: ff time})}   \\
  \midrule
  \textit{Bondi+Low-z} & Bondi-based ($\alpha=100$) & - & $\sim 4$ & - & $\mathrm{d}t_\bullet$ \\
 \textit{Sink+Low-z} & sink-based & 0.387 & $\sim 4$ & - & $\mathrm{d}t_\bullet$ \\
 \textit{Sink+FF+Low-z} & sink-based & 0.387 & $\sim 4$ & - & $\mathrm{max}[\mathrm{d}t_\bullet$,$\tau_\mathrm{FF}$] \\
  \textit{Sink+Duty+Low-z} & sink-based & 0.387  & $\sim 4$ & 25 & $\mathrm{d}t_\bullet$ \\
  \textit{Bondi+High-z} & Bondi-based ($\alpha=100$) & - & $\sim 15$ & - & $\mathrm{d}t_\bullet$ \\
  \textit{Sink+High-z} & sink-based & 0.387  & $\sim 15$ & - & d$t_\bullet$  \\
  \textit{Sink+FF+High-z} & sink-based & 0.387 & $\sim 15$ & - & $\mathrm{max}[\mathrm{d}t_\bullet$,$\tau_\mathrm{FF}$] \\
  \textit{Sink+Duty+High-z} & sink-based & 0.387 & $\sim 15$ & 25 & $\mathrm{d}t_\bullet$ \\
\bottomrule
\end{tabular}
\caption{Overview of the simulation runs. For each simulation run, the table lists the BH accretion model adopted, the value of the sink radius $R_\mathrm{sink}$, i.e. the region from which the BH accretes in the sink-based runs (see Equation~(\ref{eq: sp Mdot})), the accretion timescale $\mathrm{d}t$, and the duty cycle $\Delta t_\mathrm{AGN}$, i.e. the time interval between subsequent AGN feedback bursts. No duty cycle implies continuous injection of AGN feedback energy into the ISM, whereas in the \textit{Sink+Duty} model the AGN energy is stored for $\Delta t_\mathrm{AGN}$ and then released in a single burst.}
\label{tab: runs}
\end{table*}

With the exception of the \textit{Sink+FF} model, all sink-based setups adopt the integration timestep of the BH particle $\mathrm{d}t_\bullet$ as the timescale $\mathrm{d}t$ employed in the computation of the accretion rate in Equation~(\ref{eq: sp Mdot}). In the \textit{Sink+FF} run, we set an additional constraint on $\mathrm{d}t$, such that it cannot be shorter than the free-fall timescale $\tau_\mathrm{FF}$, i.e.
\begin{equation}
    \label{eq: ff time}
    \mathrm{d}t = \mathrm{max}[\mathrm{d}t_\bullet, \, \tau_\mathrm{FF}]; \; \; \tau_\mathrm{FF} \sim \left(\frac{\pi^2 R^3_\mathrm{sink}}{8 \mathrm{G} M_\bullet} \right)^{1/2} \, ,
\end{equation}
where $\mathrm{G}$ is the gravitational constant and $R_\mathrm{sink}$ is expressed in physical units. In this scenario, the accretion of gas onto the BH is limited by the fact that purely gravitational collapse cannot exceed the free-fall timescale within $R_\mathrm{sink}$. A caveat of this setup is that the expression for $\tau_\mathrm{FF}$ would theoretically need to account for the mass of all gas cells, stars and DM particles within the sink radius, whereas for simplicity this model only includes the mass of the BH, meaning that our estimated $\tau_\mathrm{FF}$ is likely conservative and that the actual $\mathrm{d}t$ could be lower.

Moreover, by introducing the 25 Myr duty cycle, the \textit{Sink+Duty} model will help determine whether more energetic and intermittent AGN feedback can produce results in agreement with observational constraints and empirical scaling relations. 

Together, these sink-based modelling variants are aimed at exploring how different assumptions about the effective temporal coupling between the resolved central gas reservoir and the subgrid treatment of unresolved BH accretion and AGN feedback affect the evolution of the system, while keeping the underlying galaxy formation model fixed. In particular, we stress that the SN feedback parametrisation is identical across all simulation runs, such that any differences reported in the following sections arise primarily from the adopted BH modelling choices. For a systematic study of the interplay between SN and AGN feedback within the same galaxy formation model and initial conditions, we refer the reader to \citet{koudmani2022two}.

\begin{figure*}
    \centering
    \includegraphics[width=\textwidth]{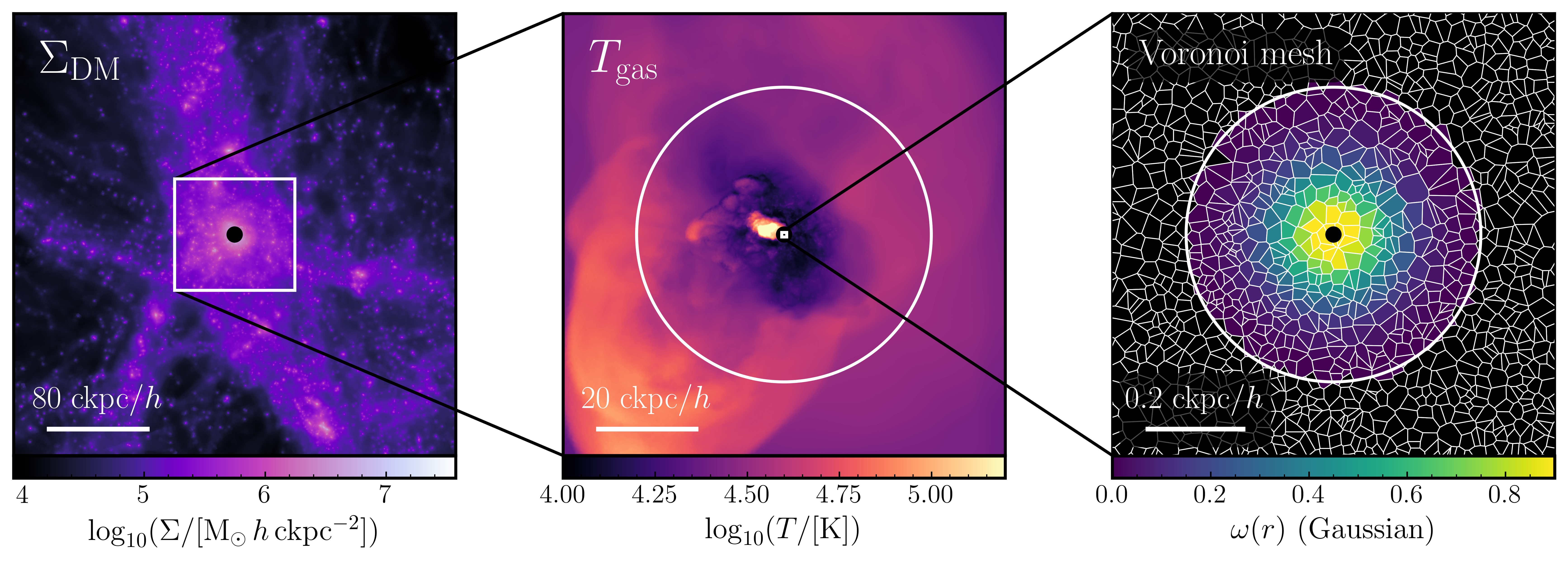}
    \caption{Zoom-in sequence of the multi-component structure of our simulated dwarf at $z \sim 3$ for the \textit{Sink+Low-z} simulation run. \textit{From left to right:} DM surface density ($\Sigma_\mathrm{DM}$) projection, gas temperature ($T_\mathrm{gas}$) projection and slice of the Voronoi mesh, all centred on the position of the BH. The projection dimensions are $(360 \; \mathrm{ckpc}/h)^3$, $(3 \times R_\mathrm{vir})^3$ and $(3 \times R_\mathrm{sink})^2$, respectively.
    In the second panel, the white circle marks the virial radius. In the third panel, the white circle indicates the sink radius, i.e. $R_\mathrm{sink} = 0.387 \; \mathrm{ckpc}$, which represents the region from which the BH can accrete gas in all sink-based runs. The black filled circle at the centre of all panels marks the position of the BH particle. Our simulated dwarf is located within a DM filament and is surrounded by numerous smaller-mass subhaloes. The middle panel shows an AGN-driven hot bubble, which expands well beyond the virial radius. The colour-coding in the last panel illustrates the sink-particle BH accretion model described in Section~\ref{sec: methods sink-particle model}, in which gas cells in the innermost regions of the galaxy contribute preferentially to BH accretion according to the Gaussian kernel $\omega(r)$ (see Equation~(\ref{eq: Gaussian kernel})).}
    \label{fig: zoom-in projections}
\end{figure*}

To qualitatively illustrate the main features of our novel sink-particle method, Figure~\ref{fig: zoom-in projections} shows a zoom-in sequence of the multi-component structure of our simulated dwarf galaxy at $z \sim 3$ for one representative run, namely the \textit{Sink+Low-z} run. From left to right, the panels display a DM surface density ($\Sigma_\mathrm{DM}$) projection, a gas temperature ($T_\mathrm{gas}$) projection and a slice of the Voronoi mesh, all centred on the position of the BH particle. The projection dimensions are $(360 \; \mathrm{ckpc}/h)^3$, $(3 \times R_\mathrm{vir})^3$ and $(3 \times R_\mathrm{sink})^2$, respectively. In the second and third panels, the regions enclosed $R_\mathrm{vir}$ and $R_\mathrm{sink}$ are indicated by white circles. Finally, the position of the central BH particle is marked with a black filled circle.

The left-hand panel reveals the filamentary structure of the DM distribution, which channels cosmic inflows towards the main subhalo. In the middle panel, AGN feedback drives a hot gas bubble, indicating efficient energy coupling to the ISM. The reasons for this behaviour will be analysed in detail in the following sections. Finally, the right-hand panel shows the structure of the Voronoi tessellation around the BH particle, demonstrating that the gas within the region enclosed by $R_\mathrm{sink}$ is well resolved by the adaptive mesh. To visually illustrate the characteristics of our novel sink-particle method, the Voronoi cells are colour-coded based on the value of the Gaussian kernel $\omega(r)$ evaluated at the each cell's position. This kernel, defined in Equation~(\ref{eq: Gaussian kernel}), enters the computation of the sink-based accretion rate by means of Equations~(\ref{eq: sp Mdot}) and~(\ref{eq: Mdrain}), such that gas cells closest to the BH contribute preferentially to accretion.  


\section{Results}
\label{sec: results}


\subsection{Cosmic evolution of the simulation runs}
\label{sec: cosmic evolution}


\subsubsection{Large-scale gas structure and thermal properties in Bondi-based vs sink-based setups}
\label{sec: projections}

\begin{figure*}
\centering
\includegraphics[width= \linewidth]{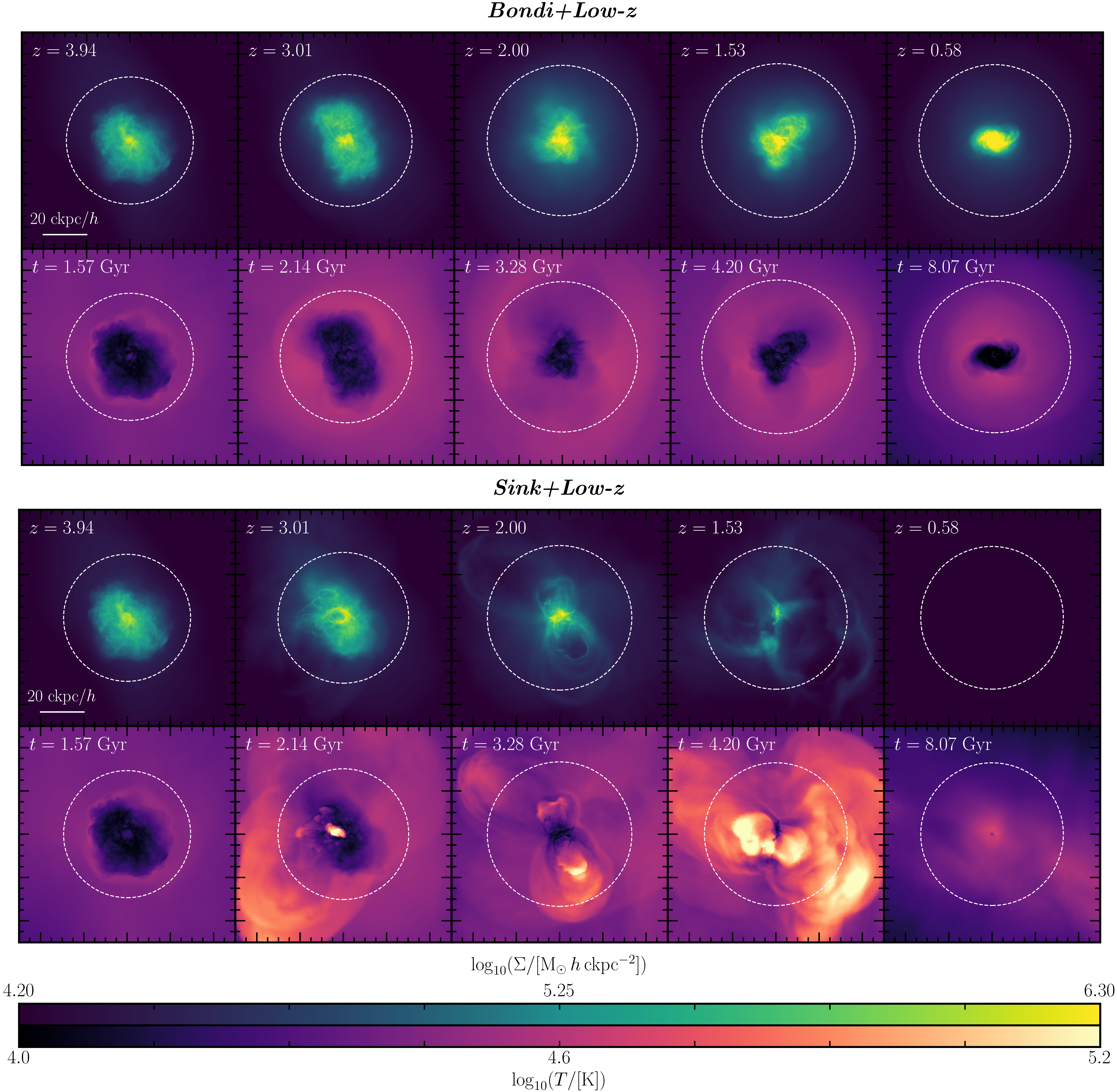}
\caption{Large-scale surface density (upper panels) and temperature (lower panels) 2D projections for two representative runs, namely the \textit{Bondi+Low-z} run and the \textit{Sink+Low-z} run. The redshifts $z$ and times $t$ of the selected snapshots are indicated in the upper left-hand corner of the upper and lower panels, respectively. The projection dimensions are fixed at $(100 \; \mathrm{ckpc}/h)^3$. The region delimited by the virial radius is represented as a white dashed circle. Note that the virial radii for the selected redshifts approximately correspond to $R_\mathrm{vir} \sim 9, 11,\, 17,\, 21,\, 33 $ kpc. In the \textit{Sink+Low-z} run, the AGN is able to heat and expel gas well beyond the virial radius, suppressing cosmic inflows.}
\label{fig: projections}
\end{figure*}

We begin our analysis by presenting large-scale gas surface density and temperature projections for two representative simulation runs, namely the \textit{Bondi+Low-z} and the \textit{Sink+Low-z} runs. The aim is to provide a first qualitative overview of the fundamental differences between our Bondi-based and sink-based models. To this end, Figure~\ref{fig: projections} shows gas surface density $\Sigma$ (upper panels) and temperature $T$ (lower panels) projections, for each selected run. In order to show the evolution of the dwarf across cosmic time, we select five representative redshifts, which are indicated in the upper left-hand corner of each panel. The projection dimensions are $(100 \; \mathrm{ckpc}/h)^3$ for all panels, and the region delimited by the virial radius $R_\mathrm{vir}$ is marked with a white dashed circle. 

In the \textit{Bondi+Low-z} run, the galaxy retains a dense central gas reservoir that becomes increasingly concentrated with cosmic time, as shown by the progressive increase in the surface density of gas in the innermost regions (upper panels). Throughout its evolution, the dwarf halo is continuously supplied with fresh gas by cosmic inflows at the scale of the virial radius, providing fuel that can ultimately reach and feed the central BH. Notably, neither the Bondi-based AGN feedback nor the more moderate SN feedback is sufficiently strong to suppress these streams by driving large-scale outflows, resulting in sustained BH accretion at all epochs. The temperature projections (lower panels) complement this picture. At early times ($z=3.94$), the central region of the dwarf hosts only moderately heated gas, with no significant hot component reaching $T \sim 10^5$ K. This region becomes increasingly concentrated with cosmic time and, by $z=0.58$, even the innermost gas has reached temperatures near the cooling floor. The sustained low temperatures of the central gas reservoir demonstrate that our model for AGN feedback is inefficient within the \textit{Bondi+Low-z} setup, as the limited AGN feedback activity is unable to drive large-scale heating and outflows.   

On the other hand, the \textit{Sink+Low-z} run displays powerful, hot AGN-driven outflows already by $z\sim3$, which carve large-scale cavities of diffuse gas that extend to $\sim1.5 \times R_\mathrm{vir}$ (upper panels). By $z=2$, these outflows have developed into under-dense bipolar lobes. It is interesting to note how, although AGN feedback is injected isotropically into the ISM, the outflows can still evolve into bipolar structures. This is because AGN-driven outflows follow the path of least resistance, propagating preferentially along directions of lower ambient gas density due to clumpy, anisotropic surroundings \citep[see e.g.][for thorough discussions]{ costa2014feedback,bourne2015resolution}. The strength of AGN feedback reaches its peak around $z=1.53$, where the outflows have expanded well beyond the virial radius of the dwarf. By $z=0.58$, the central gas reservoir has been almost completely depleted, leaving the dwarf in a quenched state and preventing further BH accretion. The gas temperature projections (lower panels) highlight the thermal energy injection by the AGN on small scales, as hot gas bubbles with $T \gtrsim 10^5$ K are launched around $z=3.01$, expand and remain buoyant. These cavities significantly heat the gas at scales beyond the virial radius, preventing fresh inflows of cold gas from the cosmic web. By $z=0.58$, the innermost region is dominated by hot, diffuse gas. Ultimately, the BH accretion model within the \textit{Sink+Low-z} setup generates efficient feedback episodes that couple to the gas on scales well beyond the virial radius of the dwarf, driving large-scale hot outflows at early times, which are able to maintain the dwarf in a quenched state. This fundamental distinction between models persists regardless of the specific seeding implementation, and it will be discussed in greater detail in Section~\ref{sec: co-evolution}.  


\subsubsection{Co-evolution of black hole and host galaxy properties across cosmic time}
\label{sec: co-evolution}

\begin{figure*}
    \centering
    \includegraphics[width=\textwidth]{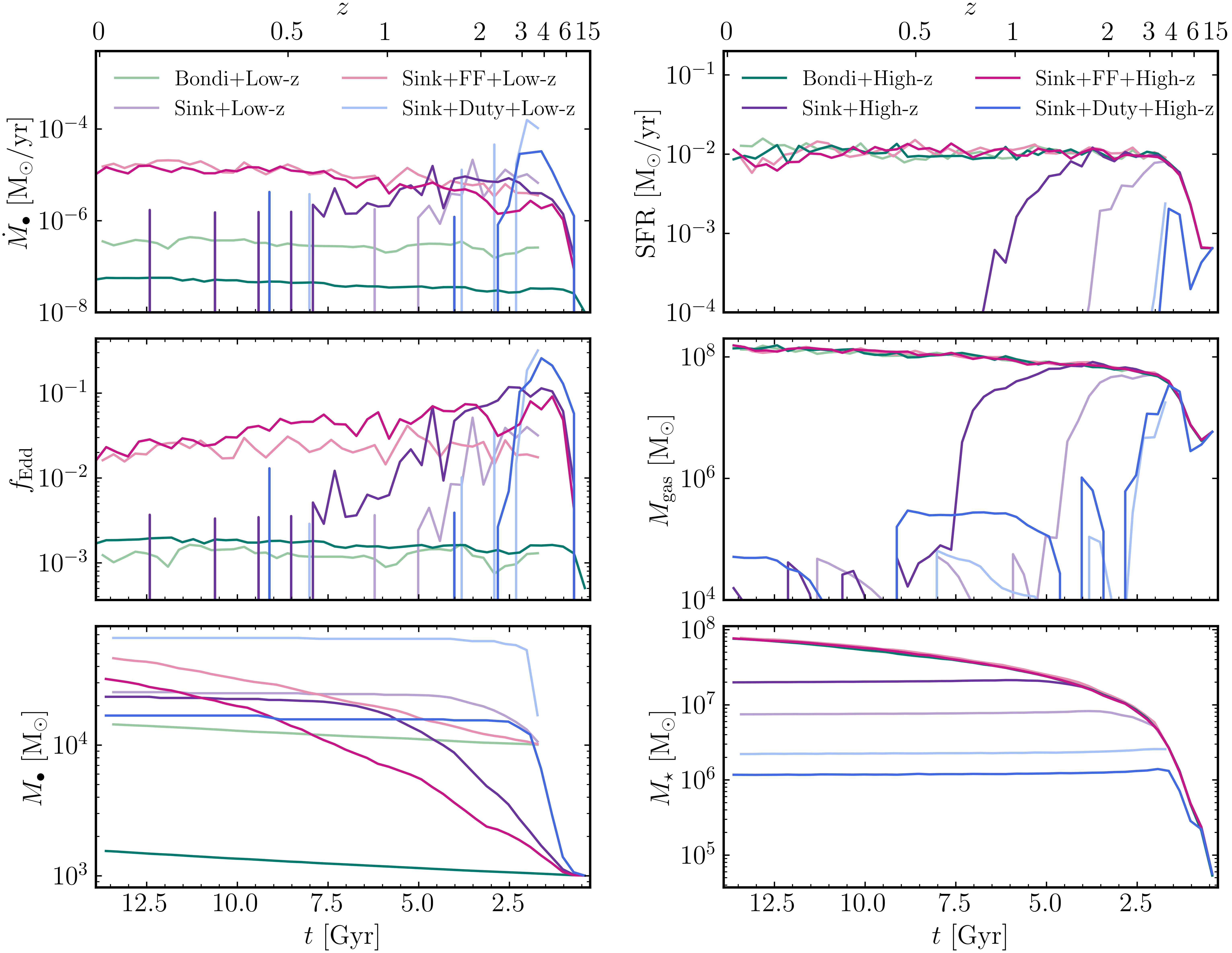}
    \caption{Evolution of BH and host galaxy properties as a function of cosmic time, for all simulation runs. \textit{Left-hand panels}: From top to bottom, the panels show the BH accretion rate $\dot{M}_\bullet$, the Eddington fraction $f_\mathrm{Edd}$ and the BH mass $M_\bullet$ against cosmic time. \textit{Right-hand panels}: From top to bottom, the panels show the SFR, the integrated gas mass $M_\mathrm{gas}$ and the integrated stellar mass $M_\star$, all calculated within twice the stellar half-mass radius, against cosmic time. All quantities are binned over $\Delta t = 0.3 \, \mathrm{Gyr}$ to even out small-scale fluctuations and emphasise overall trends. With the exception of the \textit{Sink+FF} runs, other sink-based runs are able to drive gas out of the galaxy at a redshift that depends on the specific parametrisation. After being depleted, the gas reservoir is temporarily and partially replenished before being depleted again by the next feedback burst. Efficient AGN runs are thus able to inhibit the star formation activity of the dwarf, which remains quenched at low redshifts. }
    \label{fig: cosmic evolution}
\end{figure*}

In this section, we focus on the co-evolution (or lack thereof) of BH accretion and host galaxy properties across cosmic time for all simulation runs, as illustrated in Figure~\ref{fig: cosmic evolution}. The left-hand panels display, from top to bottom, the evolution of the BH accretion rate, the Eddington fraction $f_\mathrm{Edd} = \dot{M}_\mathrm{acc}/ \dot{M}_\mathrm{Edd}$\footnote{The Eddington accretion rate is computed as $\dot{M}_\mathrm{Edd}=\frac{4\pi\mathrm{G}\mathrm{m}_\mathrm{p}}{\epsilon_\mathrm{r} \sigma_\mathrm{T}\mathrm{c}} M_\bullet$, where $\mathrm{G}$ is the gravitational constant, $\mathrm{m}_\mathrm{p}$ is the proton mass, $\epsilon_\mathrm{r}$ is the radiative efficiency, $\sigma_\mathrm{T}$ is the Thomson cross section and c is the speed of light.} and the BH mass $M_\bullet$. The right-hand panels display, from top to bottom, the evolution of the SFR, the total gas mass $M_\mathrm{gas}$ and the total stellar mass $M_\star$, all computed within twice the stellar half-mass radius. 

Focusing on the Bondi-based runs, we first note that these generally display significantly lower BH accretion rate levels (upper left-hand panel) and Eddington fractions (middle left-hand panel) compared to any sink-based run\footnote{The only exception occurs at very high redshifts ($z \gtrsim 7$) in the \textit{Bondi+High-z} run, where an initial offset of the BH from the centre of the subhalo causes it to accrete from an off-centre gas-rich clump.}. Moreover, we note that the Bondi-based accretion rate (computed via Equation~(\ref{eq: Bondi accretion})) is characterised by a heavily self-regulatory behaviour due to its proportionality to the gas density and, critically, inverse proportionality to the cube of the sound speed, which does not enter the sink-based models. This further explains why the Bondi-based runs result in low but sustained BH accretion rates across cosmic time, as the central region from which the BH accretes gas is kept rarified by moderate AGN activity. Since BH accretion is very inefficient at early times ($z>4$) in the \textit{Bondi+High-z} run due to its quadratic dependence on $M_\bullet$, the BH is not able to catch up with the BH seed mass of the \textit{Bondi+Low-z} run, as shown in the lower left-hand panel. Consequently, the \textit{Bondi+High-z} run exhibits systematically lower values of $\dot{M}_\bullet$ compared to its low-redshift counterpart. From this analysis, it appears clear that the ability to grow BHs using Bondi-based models is heavily sensitive to the choice of the initial BH seed mass, whereas this dependence is significantly milder for the sink-based runs (lower left-hand panel). 

In terms of star formation activity, the Bondi-based runs result in no appreciable impact on the SFR (upper right-hand panel) and gas reservoir (middle right-hand panel) of the dwarf, due to AGN feedback being very inefficient. Indeed, the SFR for the \textit{Bondi+Low-z} run is consistent with the no-AGN simulations from \citet{koudmani2022two}, demonstrating that the AGN does not have an appreciable impact on its host within the Bondi-based setups. Within these runs, the gas content mildly increases with cosmic time owing to gas inflows from the cosmic web. The ability of these runs to maintain a significant gas reservoir within the virial radius leads to approximately constant SFRs across cosmic time, consequently yielding a significantly higher stellar mass budget (lower right-hand panel) at $z=0$ compared to most sink-based runs. Therefore, this analysis suggests that the limitations of Bondi-based models might be too severe for exploring high BH accretion rates and efficient AGN feedback within dwarf galaxies, even after introducing the boost parameter $\alpha$. A caveat in this reasoning is that $\alpha$ could be arbitrarily increased to even higher values, leading to efficient BH accretion with Bondi-based models \citep[e.g. see][]{koudmani2022two}. However, the value of $\alpha$ is very uncertain due to the various unresolved (and poorly-understood) physical processes that it attempts to account for at low numerical cost, calling for more realistic models for BH and ISM physics in cosmological simulations (for a more detailed discussion, see Section~\ref{sec: discussion}).  

We now turn our attention to the sink-based runs, where the behaviour of BH accretion and AGN feedback activity across all diagnostics varies according to the specific modelling choices. Firstly, we note that the \textit{Sink+FF} runs are the only sink-based runs that are able to sustain efficient BH growth throughout cosmic time, showing a consistent $\sim 2$ dex difference in $\dot{M}_\bullet$ compared to the Bondi-based runs. This can be explained by considering that, similarly to the Bondi-based runs, the \textit{Sink+FF} runs are characterised by a strongly self-regulatory nature, albeit with a milder dependence on the BH mass, i.e. $\dot{M}_\mathrm{acc} \propto M_\bullet^{1/2}$ from Equation~(\ref{eq: ff time}). Additionally, it should be noted that, contrary to the Bondi-based runs, the \textit{Sink+FF} runs display a rather bursty BH accretion activity on timescales shorter than the $\Delta t = 0.3 \; \mathrm{Gyr}$ binning timescale used in our analysis. This behaviour is therefore not captured in Figure~\ref{fig: cosmic evolution}, but it is an indicator of the stronger impact of AGN feedback on the central gas reservoir in the \textit{Sink+FF} runs relative to the Bondi-based runs. It is also interesting to observe how the \textit{Sink+FF} runs display similar trends to the Bondi-based runs across all diagnostics regarding the effects of AGN feedback on the properties of the host galaxy (right-hand panels), despite showing considerable differences in their BH accretion properties (left-hand panels). This indicates that AGN feedback is inefficient in these models, resulting in no significant impact on the gas content and star formation activity of the dwarf. This can be explained once again by considering that these setups are characterised by a strongly self-regulatory nature on small scales, which prevents AGN feedback from affecting the baryon cycle of the dwarf on large scales. 

Furthermore, the key difference between the \textit{Sink+FF} runs and the other sink-based runs is that the latter are all characterised by high and sustained BH accretion rates at high redshift ($z \gtrsim 2$), followed by a dramatic decline at a redshift that depends on the specific parametrisation, with only sporadic, short-lived bursts at lower redshifts, triggered by transient gas inflows (upper left-hand panel). This is an indicator of powerful and ejective AGN feedback, which is able to expel gas from the dwarf, drastically halting its star formation activity. Note that the gas reservoir is able to partially and temporarily recover in-between subsequent feedback bursts (middle right-hand panel), although these bursts are still sufficient to maintain the galaxy in a quenched state by ejecting and heating the gas (upper right-hand panel). Additionally, note that the stellar mass of these runs slightly decreases once the galaxy is quenched, as a result of stellar evolution and mass return.

At very high redshifts ($z\gtrsim 3$), the \textit{Sink+Duty} runs display the most efficient BH accretion activity across all simulation runs\footnote{Note that the $\Delta t = 0.3 \; \mathrm{Gyr}$ binning adopted in Figure~\ref{fig: cosmic evolution} gives the impression that the BH mass in the \textit{Sink+Duty+Low-z} starts from a higher value (lower left-hand panel) This is, however, only a consequence of the temporal binning, as the BH grows significantly on a timescale shorter than the bin width, causing the early rapid growth to be averaged into the first time bin.}. This is due to the intermittent nature of AGN feedback induced by the duty cycle, which allows the gas reservoir to be replenished in between subsequent feedback injections, fueling further BH accretion activity. Therefore, the AGN duty cycle employed in this model successfully reduces overcooling and results in considerably greater impact on the ISM relative to all other models. This is because the AGN feedback energy is accumulated over $\Delta t_\mathrm{AGN}$, so that a larger amount of energy is injected in one burst into the same spatial region as in the other sink-based runs. Within this setup, the impact of the AGN on the dwarf is so significant that star formation activity is already halted by $z\sim 3$.

It is also interesting to assess the sensitivity of each BH accretion model to the choice of BH seeding configuration, i.e. the \textit{+Low-z} versus \textit{+High-z} setups. We first note that the two Bondi-based setups are fundamentally different in terms of BH properties across cosmic time. In these runs, BH growth is negligible due to inefficient accretion, such that the final BH mass at $z=0$ is essentially set by the initial seeding mass. In contrast, for a given BH accretion model, all sink-based runs display qualitatively similar trends and final outcomes across all diagnostics, largely independent of the seeding prescription. However, there are two notable exceptions to this behaviour. First, the \textit{Sink+High-z} run is able to sustain BH accretion for considerably longer than the \textit{Sink+Low-z} run, until the BH accretion rate $\dot{M}_\bullet$ abruptly drops at $z \sim 0.6$. This is likely due to the \textit{Sink+High-z} run experiencing a more gentle build-up of AGN feedback from an initially lighter BH, making it easier for the system to recover in between subsequent feedback bursts and support BH accretion for longer.   Second, in the \textit{Sink+Duty+Low-z} run the final BH mass is significantly higher than in its high-redshift counterpart, as gas available for accretion is expelled only after a brief initial phase of rapid BH growth. 

We have further verified that our qualitative conclusions are robust to the choice of sink radius, within the range of sensitivity accessible at our resolution. This was done by running two additional \textit{High-z} seeding simulations in which $R_\mathrm{sink}$ is halved and doubled. In both cases, BHs grow efficiently and drive substantial feedback well before $z=0$, in contrast to the persistently suppressed growth of the Bondi-based runs. The main effect of varying $R_\mathrm{sink}$ is to shift the onset of efficient BH growth and its associated feedback to earlier (larger $R_\mathrm{sink}$) or later (smaller $R_\mathrm{sink}$) times, with all runs converging to a similar quenched state at late times.

Additionally, to isolate the role of AGN feedback in regulating BH growth and the central gas reservoir, we ran an additional \textit{High-z} seeding simulation in which AGN feedback is switched off and accretion proceeds as in the \textit{Sink+High-z} run. In this case, the BH grows unregulated to $M_\bullet \gtrsim 10^8 \, \mathrm{M}_\odot$ by $z=0$, exceeding the stellar mass of its host, clearly inconsistent with observations of local dwarfs. While runaway accretion still depletes the central gas reservoir, the galaxy nevertheless retains substantially more gas than in the \textit{Sink+High-z} run, indicating that gas consumption by accretion alone cannot reproduce the degree of gas depletion driven by AGN feedback within our modelling framework. Therefore, once accretion is efficient, AGN feedback plays a key role in determining the final BH masses and central gas content in our simulations, consistent with the findings of \citet{koudmani2022two}.

In essence, from our analysis it emerges that efficient BH growth at high redshift ($z \gtrsim 2$) is crucial for early AGN activity to heat and expel gas beyond the virial radius, ultimately leaving the host galaxy in a quenched state. Additionally, note that in all simulation runs accretion is feedback-limited rather than Eddington-limited, meaning that the accretion rates never reach the Eddington limit (i.e. $f_\mathrm{Edd} < 1$), as AGN feedback efficiently regulates gas availability in the accretion region. As we only allow the BH to accrete from the central regions of the dwarf by limiting the BH smoothing length, this analysis further suggests that enough gas is likely able to reach these regions to fuel efficient BH growth and impactful AGN feedback activity. 

\begin{figure}
\centering
\includegraphics[width=\linewidth]{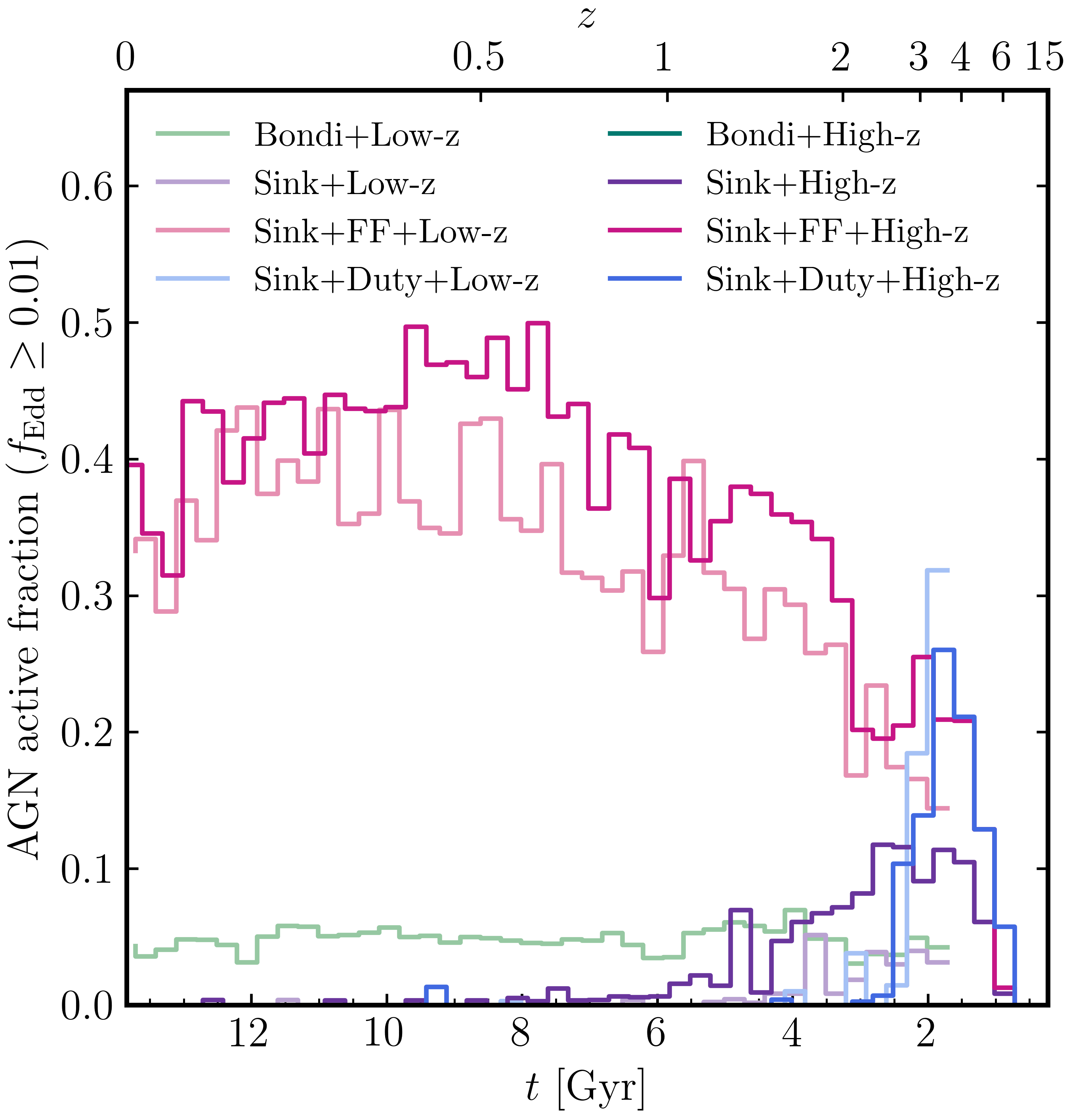}
\caption{AGN active fraction as a function of cosmic time $t$, for all simulation runs. The AGN active fraction is defined as the fraction of each time bin of width $\Delta t = 0.3 \; \mathrm{Gyr}$ during which the Eddington ratio satisfies $f_\mathrm{Edd} \geq 0.01$. Among the sink-based runs, the \textit{Sink+FF} runs generally display the highest active fraction -- especially at low redshifts -- as a result of the ability of the BH to sustain self-regulated growth across the entire redshift evolution. Other sink-based runs also display significant AGN active fraction at high redshift, while due to efficient quenching of their host, BHs in these runs are largely dormant at low redshifts.}
\label{fig: AGN actve fraction}
\end{figure}

It should be noted that the temporal binning adopted in Figure~\ref{fig: cosmic evolution} naturally smooths over rapid fluctuations in BH accretion and AGN feedback. While this binning choice is deliberately made to highlight long-term trends, we also wish to emphasize the underlying bursty behaviour of these processes. To this end, Figure~\ref{fig: AGN actve fraction} shows the AGN active fraction as a function of time $t$, for all simulation runs. The AGN active fraction is hereby defined as the portion of each $\Delta t = 0.3 \; \mathrm{Gyr}$ bin during which the Eddington ratio $f_\mathrm{Edd}$ meets or exceeds $0.01$. 

From this analysis, it first emerges that the \textit{Bondi+High-z} run never reaches $f_\mathrm{Edd} = 0.01$ due to the low BH seed mass, whereas the \textit{Bondi+Low-z} exhibits consistently low AGN active fractions throughout cosmic time, rarely exceeding the threshold. Among the sink-based models, the \textit{Sink+FF} runs display the highest AGN active fractions, reaching a peak of $\sim 0.5$ at $z\sim 0.7$. This confirms the bursty but sustained nature of BH accretion in these runs, which leads to mild AGN feedback episodes that regulate BH accretion in a cyclic fashion. Finally, the remaining sink-based runs display high AGN active fractions at early times $z \gtrsim 2$, followed by a rapid decline. This is due to the AGN-driven depletion of the central gas reservoir, which results in a mostly dormant AGN phase at later times, interrupted only by occasional, short-lived feedback bursts. 

Crucially, efficient AGN activity at high redshift is increasingly probed by recent \textit{JWST} observations of luminous low-mass AGN in the early Universe, highlighting this epoch as particularly promising for studying early BH growth and constraining the AGN luminosity function \citep[e.g.][]{maiolino2024jades, geris2026jades, hickox2025exploring}. Although our simulated systems do not represent direct counterparts of these sources due to their lower masses, our results indicate that early BH growth and AGN feedback remain highly sensitive to the subgrid BH seeding and accretion prescriptions in low-mass galaxies. The different outcomes explored here therefore provide bracketing cases that quantify how modelling choices can systematically affect growth histories and feedback efficiencies, highlighting the associated uncertainty that must be considered when interpreting current and future \textit{JWST} observations.


\subsection{Validation of dynamical mass estimators against simulated dwarf galaxies}
\label{sec: Mdyn estimators}

Galaxy masses play a central role in testing theories of cosmological structure formation \citep[see e.g.][for a review]{courteau2014galaxy}. Among the most robust empirical probes of galaxy mass are dynamical measurements, such as the line-of-sight (LOS) velocity dispersion of the stellar component, $\sigma_\star$, and the \textit{dynamical mass}, $M_\mathrm{dyn}$, which trace the depth of the gravitational potential well and are observed to correlate with the central BH mass \citep[e.g.][]{ferrarese2000fundamental, gebhardt2000relationship, kormendy2013coevolution}. Observationally, the dynamical mass is inferred as the mass required to reproduce the observed kinematics of a tracer, such as stars or gas, under the assumption of virial equilibrium \citep[e.g.][]{binney1987galactic}. In practice, direct application of the idealised scalar virial theorem is often limited by dynamically unresolved systems, departures from virial equilibrium, and environmental perturbations, e.g. inflows, outflows and tidal interactions. As a result, observational studies often rely on simplified virial estimators calibrated on simulations or well-resolved stellar kinematic data. A widely adopted approach is to use the LOS velocity dispersion, $\sigma$, of a dynamical tracer (stars or gas) to infer $M_\mathrm{dyn}$ as the total mass enclosed within an \textit{effective} radius $R_\mathrm{e}$ -- typically the projected stellar half-light radius or a closely related scale \citep[e.g.][]{walker2009universal, wolf2010accurate, errani2018systematics}. In this work, we follow previous studies that adopt a factor of two to approximate the total mass within a larger radius, i.e. $M_\mathrm{dyn} \equiv 2 \times M(< R_\mathrm{e})$ \citep[e.g.][]{cappellari2006sauron, van2022mass}.

Various \textit{JWST}-based analyses \citep[e.g.][]{ubler2023ga, maiolino2024jades} employ the virial estimator from \citet{van2022mass}, which was calibrated using dynamical models for a sample of galaxies at $z = 0.6 - 1.0$ from the LEGA-C survey, with $M_\star \gtrsim 2 \times 10^{10} \; \mathrm{M_\odot}$. This calibration reads: 
\begin{equation}
    \label{eq: vdW M_dyn}
    M_\mathrm{dyn} = K(n) K(q) \frac{\sigma^2 R_\mathrm{e}}{\mathrm{G}} \; ,
\end{equation}
where $K(n)$ and $K(q)$ are structural factors that correct for the concentration and geometry of the galaxy, with $K(n) = 8.87 - 0.831n + 0.0241n^2$ \citep[with $n$ the Sérsic index, from][]{bertin2002weak, cappellari2006sauron} and $K(q) = [0.87 + 0.38 \mathrm{e}^{-3.78(1-q)} ]^2$ \citep[with $q$ the projected axis ratio, from][]{van2022mass}. Here, $R_\mathrm{e}$ is the effective radius, i.e. the projected stellar half-light radius, and $\sigma$ is the LOS velocity dispersion of the dynamical tracer within $R_\mathrm{e}$. 

In the dwarf-galaxy regime, where stellar masses are significantly lower and systems are often dispersion-supported, similar virial-based approached have been developed. The most widely employed are those from \citet{walker2009universal} and \citet{wolf2010accurate}, which have been tested on cosmological simulations of dwarf galaxies, reproducing the true enclosed masses within $\sim 20-25 \%$ \citep[e.g.][]{campbell2017knowing, gonzalez2017dwarf}. In this section, we focus on the \citet{walker2009universal} estimator, as the definition of $R_\mathrm{e}$ is broadly  consistent with the one from \citet{van2022mass}. The \citet{walker2009universal} estimator was calibrated using dynamical models for the eight brightest dwarf spheroidal galaxies in the Local Group. This calibration reads:
\begin{equation}
    \label{eq: Walker M_dyn}
    M_\mathrm{dyn} = 5 \frac{\sigma^2 R_\mathrm{e}}{\mathrm{G}} \; ,
\end{equation}
where $\sigma$ is now measured over the entire galaxy.  

at high redshift, direct measurements of the stellar velocity dispersion, $\sigma_\star$, are often limited by low signal-to-noise ratios on the continuum. In contrast, the gas velocity dispersion, $\sigma_\mathrm{gas}$, can be estimated from the widths of strong nebular emission lines (e.g., H$\alpha$, [OIII]5007), providing a practical proxy when $\sigma_\star$ is unavailable. Physically, the integrated emission-line width captures both random gas motions and projected rotational shear, probing the same underlying gravitational potential as $\sigma_\star$, which naturally leads to a correlation between the two \citep[e.g.][]{greene2005comparison, oh2022sami, bezanson20181d}. However, since gas is a collisional fluid, its dynamics are also dictated by non-gravitational processes, including cooling, heating, turbulence, shocks, magnetic fields and radiation pressure \citep[see e.g.][for a thorough discussion]{kormendy2013coevolution}, which can broaden emission lines and bias dynamical mass estimates. In particular, strong AGN feedback can drive fast, large-scale outflows that decouple the gas and stellar kinematics \citep[e.g.][]{penny2018sdss, koudmani2019fast}. Thus, one should proceed with great caution and on a case-by-case basis when using $\sigma_\mathrm{gas}$ as a proxy for $\sigma_\star$, carefully evaluating whether the gravitational processes that govern the stellar kinematics are the dominant source of gas kinematics.

In this section, we compare dynamical masses estimated via Equation~(\ref{eq: vdW M_dyn}) and Equation~(\ref{eq: Walker M_dyn}) with the true values of $2 \times M(< R_\mathrm{e})$, computed via direct summation from our simulations. For each estimator, we further examine the impact of adopting either the LOS stellar velocity dispersion, i.e. $\sigma \equiv \sigma_\star$, or the LOS gas velocity dispersion, i.e. $\sigma \equiv \sigma_\mathrm{gas}$, in the dynamical mass computation. This allows us to assess the validity of these estimators in our simulated dwarfs, as well as to approximately quantify the effect of using the gas velocity dispersion as a proxy for the stellar velocity dispersion, particularly in systems subject to strong AGN feedback. It is crucial to note that the stellar masses of our simulated dwarfs never exceed $M_\star \sim 1 \times  10^{8} \; \mathrm{M_\odot}$ (see lower right-hand panel in Figure~\ref{fig: cosmic evolution}), such that these systems are placed well outside the stellar mass range over which the \citet{van2022mass} estimator was calibrated, thus requiring significant extrapolation. 

For each snapshot in a simulation, we compute the effective radius $R_\mathrm{e}$ and Sérsic index $n$ by fitting a Sérsic profile to circularly-binned projected stellar surface mass density profiles\footnote{In observational studies, Sérsic fits are performed on stellar surface brightness profiles. In our simulations, we implicitly assume that stellar mass directly traces stellar light by taking the surface mass density profile as a proxy for the surface brightness profile.} in 100 randomly-generated lines of sight, similar to previous simulation work \citep[e.g.][]{gonzalez2017dwarf}.
The axis ratio $q$ is calculated from the eigenvalues of the mass-weighted 2D inertia tensor of the projected stellar positions. Both the Sérsic fits and the computation of $q$ are carried out within a projected radius equal to $4 \times R_\mathrm{hm,\star}$, where $R_\mathrm{hm,\star}$ is the 3D stellar half-mass radius. As shown in Figure~\ref{fig: q, n, R_e violins} for the \textit{Sink+Low-z} run, the median projected axis ratios of our simulated dwarf are typically $q \gtrsim 0.7$ across a range of redshifts, indicating that the stellar distributions are nearly spherical. This suggests that circularly-binned profiles provide a robust approximation for the Sérsic fits, in line with the findings of \citet{gonzalez2017dwarf}. The resulting distributions of $R_\mathrm{e}$, $q$ and $n$ for four representative redshifts within the \textit{Sink+Low-z} run are reported in Figure~\ref{fig: q, n, R_e violins}. 

For each projection, we then measure the mass-weighted LOS stellar and gas velocity dispersions, $\sigma_\star$ and $\sigma_\mathrm{gas}$, by selecting all particles within a projected radius either equal to $R_\mathrm{e}$ (for the \citet{van2022mass} estimator) or $4 \times R_\mathrm{e}$ (for the \citet{walker2009universal} estimator). The latter choice is motivated by two factors: for  $r \gtrsim 4 \times R_\mathrm{e}$ the number of stellar particles becomes negligible compared to the inner regions, and for well-resolved dwarf galaxies most available kinematic data extend out to $r \sim 5 \times R_\mathrm{e}$ \citep[e.g.][]{walker2009universal}.
These quantities are then used in Equation~(\ref{eq: vdW M_dyn}) and Equation~(\ref{eq: Walker M_dyn}) to compute the virial dynamical masses, $M_\mathrm{dyn}^\mathrm{obs,\star}$ (i.e. setting $\sigma \equiv \sigma_\star$) and $M_\mathrm{dyn}^\mathrm{obs,gas}$ (i.e. setting $\sigma \equiv \sigma_\mathrm{gas}$), for each estimator. The true dynamical mass, $M_\mathrm{dyn}^\mathrm{sim}$, is instead computed by summing the masses of all collisionless particles (DM, stars and BH) and gas cells within a sphere of radius $R_\mathrm{e}$ and multiplying by two.

\begin{figure*}
\centering
\includegraphics[width=0.8\linewidth]{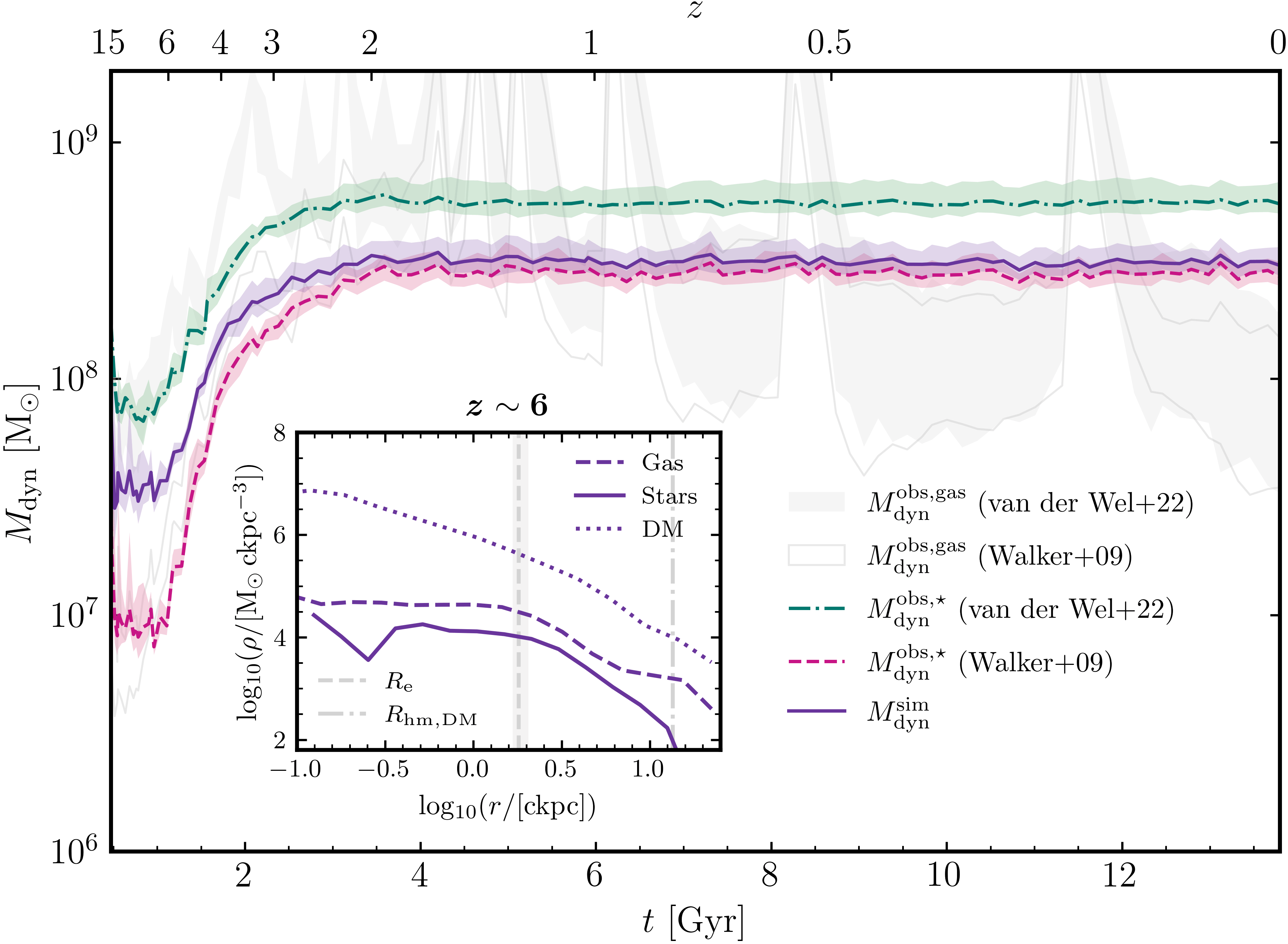}
\caption{
\textit{Main panel:} Evolution of the dynamical mass as a function of cosmic time for the \textit{Sink+Low-z} setup. The dynamical mass is defined as $M_\mathrm{dyn} \equiv 2 \times M(<R_\mathrm{e})$, where $R_\mathrm{e}$ is the 2D projected stellar half-mass radius. At each snapshot, we perform 100 random projections, measure the projected axis ratio $q$, and carry out Sérsic fits to determine the Sérsic index $n$ and $R_\mathrm{e}$. We then compare the true dynamical mass measured directly from the simulations ($M_\mathrm{dyn}^\mathrm{sim}$; purple solid line) with the virial estimators from \citet{van2022mass} (green dash-dotted line) and \citet{walker2009universal} (pink dashed line). We label these inferred masses as $M_\mathrm{dyn}^\mathrm{obs,\star}$ as they were derived by setting $\sigma \equiv \sigma_\star$ (see Equation~(\ref{eq: vdW M_dyn}) and Equation~(\ref{eq: Walker M_dyn})). Shaded regions around each line correspond to the 16th–84th percentile range, capturing the scatter across the projections. For comparison, we also show the 16th–84th percentile ranges obtained by setting $\sigma \equiv \sigma_\mathrm{gas}$ in the same estimators, which we denote as $M_\mathrm{dyn}^\mathrm{obs, gas}$. These are indicated as a light-grey shaded band for the \citet{van2022mass} estimator and as an unfilled band bounded by light-grey lines for the \citet{walker2009universal} estimator.
\textit{Inset:} Radial profiles of the 3D mass density $\rho$ for DM (dotted line), stars (solid line), and gas (dashed line) within the virial radius at $z \sim 6$. The median effective radius ($R_\mathrm{e}$) and the 3D DM half-mass radius ($R_\mathrm{hm,DM}$) are marked with vertical grey dashed and dash-dotted lines, respectively. The 16th and 84th percentiles for $R_\mathrm{e}$ are shown as a vertical grey shaded band. As our simulated dwarf falls well outside the stellar mass range of the galaxies used for calibrating the \citet{van2022mass} estimator, extrapolating this calibration to such low-mass, DM-dominated systems is likely invalid and leads to a systematic overestimation of $M_\mathrm{dyn}$. The \citet{walker2009universal} estimator recovers the true dynamical mass to within $\sim 10-15 \%$, albeit only for $z \lesssim 2$.}
\label{fig: M_dyn vs t}
\end{figure*}

The results of this pipeline are illustrated in the main plot within Figure~\ref{fig: M_dyn vs t}, which shows the evolution of different estimates of the dynamical mass across cosmic time for one representative run, namely the \textit{Sink+Low-z} run. The purple solid line represents the median direct measurement of $M_\mathrm{dyn}^\mathrm{sim}$ from the simulations, while the pink dashed line and the green dash-dotted line represent the median values of $M_\mathrm{dyn}^\mathrm{obs,\star}$ using the \citet{walker2009universal} and \citet{van2022mass} estimators, respectively. Shaded bands around each line show the corresponding 16th and 84th percentiles, reflecting the scatter across all random projections performed at each snapshot. To assess the effects of using the gas velocity dispersion, Figure~(\ref{fig: M_dyn vs t}) also shows, for both estimators, the 16th-84th percentile ranges of the dynamical masses inferred by setting $\sigma \equiv \sigma_\mathrm{gas}$ in Equation~(\ref{eq: vdW M_dyn}) and Equation~(\ref{eq: Walker M_dyn}), which we label as $M_\mathrm{dyn}^\mathrm{obs, gas}$. These gas-based ranges are shown in light grey, with a filled band for the \citet{van2022mass} estimator and an unfilled band bounded by light-grey lines for the \citet{walker2009universal} estimator. Additionally, the inset plot displays the radial profiles of the 3D mass density $\rho$ within the virial radius for DM (purple dotted line), stars (purple solid line) and gas (purple dashed line) at $z\sim 6$. The median effective radius from the Sérsic fit, $R_\mathrm{e}$, and the 3D DM half-mass radius, $R_\mathrm{hm,DM}$, are marked with vertical grey dashed and dash-dotted lines, respectively. This comparison highlights the relative contribution of each component across different radii, showing that our simulated dwarf is DM-dominated at all radii. Further details on how the profiles are constructed, as well as their evolution across cosmic time, are provided in Appendix~\ref{sec: radial density profiles appendix}. 

We first focus on the case in which the velocity dispersion entering the dynamical mass estimators is taken to be the LOS stellar velocity dispersion, i.e. $\sigma \equiv \sigma_\star$. At very early times ($z \gtrsim 6$), all measurements of the dynamical mass, including the true dynamical mass, show significant scatter across projections. This is a direct consequence of the fact that stellar assembly is still in its early stages, such that the stellar distribution is very irregular. As a result, attempts to fit these distributions with a smooth Sérsic profile are fundamentally unreliable, ultimately biasing the inferred value of $R_\mathrm{e}$. Projection effects further amplify these fluctuations, as a small number of stars can dominate the measured kinematics along a given LOS. Essentially, in this high-redshift regime, differences between various estimators are primarily driven not by their methodological details, but rather by the evolving morphology and kinematics of the stellar component itself. As the structure of the stellar distribution becomes increasingly regular and the system reaches virial equilibrium, at $z \lesssim 2$ the \citet{walker2009universal} estimator accurately recovers the true dynamical mass of our simulated dwarf within a factor of $10-15 \%$, when the stellar velocity dispersion is used. On the contrary, the \citet{van2022mass} estimator consistently overestimates the dynamical mass by a factor of $1.7-1.9$, indicating that one or more physical assumptions behind this estimator are violated in our simulated dwarf galaxy. Moreover, as discussed earlier in this section, the stellar mass of our galaxy is placed well outside the calibration range of the \citet{van2022mass} estimator, further exacerbating this systematic offset. Physically, one potential reason for this discrepancy is that our simulated dwarf is strongly DM-dominated at all radii and across cosmic time (see inset plot within Figure~\ref{fig: M_dyn vs t} and Figure~\ref{fig: 3D profiles DM vs stars vs gas}). The shape of the mass distribution of the DM component also differs significantly from that of the stars, such that the stellar component may not trace the underlying gravitational potential in a manner similar to the calibration sample. On the other hand, the \citet{walker2009universal} estimator, calibrated specifically on DM-dominated dwarf galaxies with comparable stellar masses at $z=0$ to our simulated system, remains reasonably accurate across a wide range of redshifts. Note that this calls into question using the \citet{van2022mass} estimator for low-mass and DM-dominated galaxies. Moreover, both estimators do not perform well at high redshift ($z \gtrsim 4$), during the initial dwarf assembly, which is the cosmic epoch now probed by \textit{JWST} observations. This behaviour is consistent across all simulation runs and throughout cosmic time, as demonstrated in Figure~\ref{fig: M_dyn vs t all sims}, and is therefore not dependent on AGN feedback. 

Furthermore, regardless of the choice of virial estimator, our results indicate that the gas velocity dispersion serves as a good proxy for the stellar velocity dispersion only when AGN feedback is inefficient (see Figure~\ref{fig: M_dyn vs t all sims}), while we find no correlation between gas and stellar kinematics in setups exhibiting efficient AGN feedback activity. For example, in the \textit{Sink+Low-z} run, $\sigma_\mathrm{gas}$ traces $\sigma_\star$ only at $z>4$, prior to BH seeding. With the onset of strong AGN feedback, the gas kinematics largely decouple from the stellar kinematics, suggesting that feedback effects are dominating over gravity in dictating the motion of the gaseous component. In these regards, it is particularly interesting to note how our estimate of $M_\mathrm{dyn}$ from $\sigma_\mathrm{gas}$ traces the behaviour and strength of AGN feedback across cosmic time, which was  analysed in Section~\ref{sec: cosmic evolution}. Additionally, despite the increasingly infrequent and bursty nature of BH accretion and AGN feedback at later cosmic times, these occasional bursts are still sufficient to sustain the kinematic decoupling of the gaseous and stellar components down to $z=0$.

An important caveat in this analysis is that we do not distinguish between different gas phases in our computation of $\sigma_\mathrm{gas}$, while observational studies typically isolate the cold, star-forming ISM when measuring gas kinematics, and in the case of \citet{juodvzbalis2025direct} find a rotationally supported gaseous disc. As a result, our $\sigma_\mathrm{gas}$ measurements include contributions from the warm and hot gas phases, which are more strongly affected by AGN feedback, potentially enhancing the kinematic decoupling. A detailed study of phase-dependent gas kinematics is beyond the scope of this work, as our simulations do not resolve the multi-phase ISM. 


\subsection{Scaling relations}
\label{sec: scaling relations}

In this section, we examine how our simulated dwarf galaxies populate key BH--host galaxy scaling relations: BH mass vs stellar mass ($M_\bullet - M_\star$), BH mass vs stellar velocity dispersion ($M_\bullet - \sigma_\star$), and BH mass vs dynamical mass ($M_\bullet - M_\mathrm{dyn}$). 


\subsubsection{$M_\bullet - M_\star$ relation}
\label{sec: M_BH vs M_star relation}

\begin{figure*}
\centering
\includegraphics[width=0.8\linewidth]{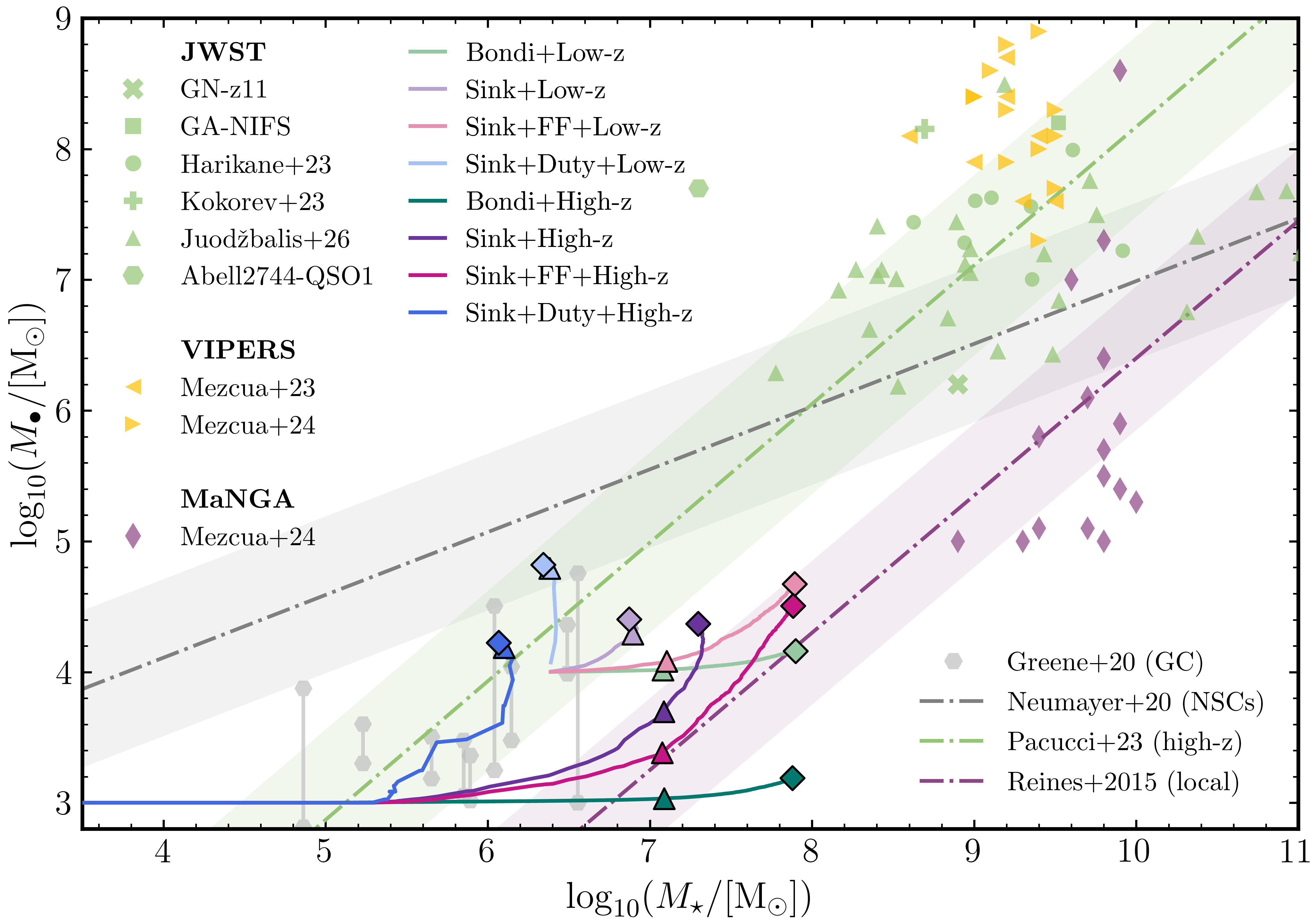}
\caption{Cosmic evolution of all simulation runs in $M_\bullet - M_\star$ phase space. $M_\star$ is computed within twice the stellar half-mass radius. The simulation tracks are represented as solid lines, with triangles and diamonds representing the $z=2$ and $z=0$ simulated data points, respectively. For comparison, we plot observational data taken at various redshifts using scatter points, as indicated in the legend. We also plot the extrapolated $M_\bullet - M_\star$ empirical scaling relations from \citet{pacucci2023jwst} (high-redshift) and \citet{reines2015relations} (local), as well as the $M_\mathrm{NSC} - M_\star$ scaling relation from \citet{neumayer2020nuclear}, which can be considered as an upper limit. The most efficient AGN runs are overall in good agreement with the extrapolated high-redshift empirical scaling relation, whereas they are overmassive compared to the extrapolated local relation, hinting at a potential link with the high-redshift overmassive BHs probed by \textit{JWST}.}
\label{fig: Mstar vs MBH}
\end{figure*}

We first focus on the cosmic evolution of all simulation runs in $M_\bullet - M_\star$ phase space, as shown in Figure~\ref{fig: Mstar vs MBH}. In order to remain broadly consistent with the majority of observational studies, $M_\star$ is hereby defined as the total stellar mass within twice the stellar half-mass radius. We note that observed stellar masses are typically derived from SED fitting and are therefore subject to systematics such as surface-brightness dimming, outshining and IMF assumptions, though a full treatment of these effects is beyond the scope of this work.

For reference, the simulated data points at $z = 2$ and  $0$ are highlighted as triangles and diamonds, respectively, for all runs. Although the $M_\bullet - M_\star$ scaling relation remains largely unconstrained in the stellar mass regime explored in this work due to intrinsic observational challenges, it is nevertheless informative to compare our results to extrapolated scaling relations and recent observational data across cosmic time. 

To highlight the evolution of the $M_\bullet - M_\star$ scaling relation across cosmic time, we plot the empirical relations from \citet{pacucci2023jwst} (high-redshift, green dash-dotted line) and \citet{reines2015relations} (local, purple dash-dotted line), along with the $1\sigma$ uncertainty on the fit parameters (shaded bands). The high-redshift relation deviates at $>3\sigma$ from the local relation, suggesting that high-redshift BHs may be overmassive by $\sim 1-2$ dex with respect to their local counterparts for a given stellar mass. However, note that this trend could be largely driven by selection bias, as high-redshift AGN samples are typically skewed towards luminous AGN relative to their host galaxies \citep[e.g.][]{li2025tip, geris2026jades}, such that currently detected overmassive \textit{JWST} BHs may not be representative of the entire population. Keeping this in mind, it is still informative to understand how such objects may have formed in the high-redshift Universe. 

Furthermore, it is important to emphasise that it remains uncertain whether extrapolating the $M_\bullet - M_\star$ scaling relation is valid for our regime of interest. With initial BH masses of $M_\bullet \sim 10^3 - 10^4 \; \mathrm{M_\odot}$, we are effectively probing the seed-mass domain, where the $M_\bullet - M_\star$ relation may flatten \citep[e.g.][]{greene2020intermediate}. Additionally, some studies suggest lower normalisations in galaxies with broad-line AGN \citep[e.g.][]{reines2015relations}. 

We also plot the scaling relation between NSC mass and stellar mass of the host ($M_\mathrm{NSC} - M_\star$) from \citet{neumayer2020nuclear} (grey dash-dotted line), which can likely be regarded as an upper limit, as the NSC mass is expected to dominate over the mass of the central BH in galaxies with stellar masses within or below $M_\star \lesssim 10^{8-10}$ M$_\odot$. Note that the data sample used in \citet{neumayer2020nuclear} includes measurements that reach stellar masses below $10^6 \; \mathrm{M}_\odot$ so that the relation is well-defined for the stellar mass range of interest in this work. 

We then plot observational data covering a wide range of redshifts ($z\sim 0-11$), as a means to show the cosmic evolution of the scatter around the (redshift-dependent) $M_\bullet - M_\star$ relation. A compilation of high-redshift ($z \sim 2-11$) \textit{JWST} observations is shown, including the sample of 21 galaxies used to derive the high-redshift relation in \citet{pacucci2023jwst}: 12 broad-line AGN at $z\sim4-11$ from JADES observations \citep[][]{maiolino2024jades}, 8 H$\alpha$ broad-line AGN at $z\sim4-7$ from CEERS and GLASS observations \citep[][green circles]{harikane2023jwst}, and one massive BH in a low-metallicity AGN at $z\sim 5.55$ from NIRSpec IFS observations \citep[][green square]{ubler2023ga}. Note that the data from \citet{maiolino2024jades} is a subset of the spectroscopic sample of Type 1 AGN from JADES spanning $1.5 < z < 9.0$ \citep[][green triangles]{juodvzbalis2026jades}, and we therefore only plot the latter. We also show two additional \textit{JWST}-NIRSpec sources: GN-z11 at $z\sim 10.6$ \citep[][green cross]{maiolino2024small}, and a broad-line AGN at $z=8.50$ from the UNCOVER Treasury survey \citep[][green `$+$' symbol]{kokorev2023uncover}. Finally, we include the recent, direct dynamical BH mass measurement in a strongly lensed Little Red Dot (LRD) at $z = 7.04$, which reveals a BH with $M_\bullet/M_\star > 2$ \citep[][green hexagon]{juodvzbalis2025direct}. 

It should be mentioned that both the BH mass and stellar mass measurements are highly uncertain in these distant systems.
In most cases BH masses are inferred from single epoch measurements, from the width and luminosity of the broad component of H$\alpha$, by using locally calibrated virial relations. The validity of these relations at high redshift has been questioned, in particular for the population of LRDs, for which it has been claimed that lines are broadened by electron scattering \citep[e.g.][]{Chang2025}. The latter claim has been disputed by other works \citep[e.g.][]{juodvzbalis2026jades,Brazzini2025}. Additionally, \citet{DEugenio25_irony} have shown that the change in BH mass is modest even when taking into account the putative electron scattering. Reassuringly, the direct BH mass measurement in a lensed LRD by \cite{juodvzbalis2025direct} has shown full consistency with the estimates provided by the virial relations. Finally, most of the AGN in the samples shown in Figure~\ref{fig: Mstar vs MBH} are not LRDs. As for the stellar masses, these are certainly also uncertain. However, the finding that Balmer breaks observed in the spectra of these objects are often associated with dense gas absorption and not stellar in origin \citep[e.g.][]{Inayoshi_Maiolino2025,ji2025blackthunder} suggests that the stellar masses inferred from SED fitting are likely upper limits, hence exacerbating the overmassive nature of these BHs.

To connect high-redshift and local observations, Figure~\ref{fig: Mstar vs MBH} also includes a sample of 12 supermassive BHs in low-mass hosts at cosmic noon ($z\sim1-3$) from the VIPERS survey \citep[][right-pointing gold triangles]{mezcua2024overmassive}, which are found to be overmassive by $\sim 2$ dex with respect to the stellar mass of their hosts, assuming that the local scaling relation holds in this redshift regime. Moreover, we plot intermediate-redshift ($z\sim 0.35-0.93$) data from the VIPERS survey \citep[][left-pointing gold triangles]{mezcua2023overmassive}, including 7 broad-line AGN powered by overmassive BHs in dwarf galaxies. We also show a sample of local broad-line AGN in dwarfs from the MaNGA survey, with BH masses in the IMBH regime \citep[][purple diamonds]{mezcua2024manga}. Finally, we include a sample of globular clusters \citep[][grey hexagons]{greene2020intermediate}. Note that the presence of BHs in globular clusters remains highly debated, and the associated BH mass measurements are subject to significant uncertainties, which we indicate by plotting error bars on $M_\bullet$.

Focusing on the cosmic evolution of the simulation runs, we can identify two main trends, based on the location of our simulated systems in $M_\bullet - M_\star$ space at $z\sim0$: \textit{efficient} AGN runs falling within the scatter of the high-redshift relation (i.e. \textit{Sink+Low-z}, \textit{Sink+Duty+Low-z} and \textit{Sink+Duty+High-z}), and \textit{inefficient} AGN runs falling within the scatter of the local relation (i.e. \textit{Bondi+Low-z}, \textit{Sink+FF+Low-z} and \textit{Sink+FF+High-z}). In essence, efficient AGN runs are characterised by substantial, early BH growth and star formation quenching, while inefficient AGN runs are characterised by moderate BH growth and sustained star formation activity across cosmic time. There is an intermediate case placed between the two relations at $z\sim0$, namely the \textit{Sink+High-z} run. The key difference between this run and other sink-based runs is not its final BH mass, but rather the ability to sustain star formation activity until later times compared to other efficient AGN runs (see upper right-hand panel in Figure~\ref{fig: cosmic evolution}), while still yielding a lower final stellar mass compared to inefficient AGN runs due to AGN-driven quenching. Additionally, there is another outlier, the \textit{Bondi+High-z} run, which does not fall within the scatter of any of the relations. Within this setup, the limitations given by the extremely inefficient Bondi-based accretion due to the low seed mass are too severe for it to reach the local scaling relation. 

In general, we find that models associated with efficient BH accretion and strong AGN feedback at early times yield BHs that are overmassive with respect to the extrapolated local scaling relation for a given stellar mass. Instead, these runs mostly fall within the scatter of the high-redshift relation. Furthermore, all simulation runs fall below the $M_\mathrm{NSC}-M_\star$ scaling relation, consistent with local observations.

It is important to emphasise that although efficient AGN runs result in overmassive BHs at $z\sim0$ compared to the local relation, this does not rule out the possibility of efficient BH accretion and AGN feedback in dwarfs, since the shape, normalisation and scatter of the local scaling relation in this mass regime have not yet been probed observationally. 

From a purely theoretical standpoint, our simulations suggest two parallel, plausible evolutionary pathways. On the one hand, efficient AGN runs naturally lead to phases in which BH growth outpaces the assembly of the host galaxy, producing overmassive BHs with respect to the extrapolated local relation, while remaining broadly consistent with the normalisation implied by recent high-redshift extrapolations. The emergence of overmassive BHs is hence best understood as a manifestation of the sensitivity of early BH–galaxy co-evolution to the regulation of gas inflow due to a number of complex physical processes, including gas angular momentum transport as well as feedback efficiency and modus operandi.
Such low-mass systems could, in principle, retain long-lived imprints of early phases of efficient BH growth, potentially appearing today as dormant, observationally elusive overmassive BHs relative to their present-day stellar mass content. Although the mass regimes differ, this phenomenology qualitatively overlaps with the regime now being probed by \textit{JWST} observations of high-redshift, low-mass AGN, raising the intriguing possibility that overmassive BHs may represent a more general manifestation of rapid early BH growth in low-mass systems across cosmic time. On the other hand, the inefficient AGN runs, initially located within the scatter of the high-redshift relation, gradually evolve towards the local relation at $z\sim 0$, due to stellar build-up dominating over BH accretion. 

We caution that there are caveats in this analysis, which will be discussed in detail in Section \ref{sec: caveats}. One major caveat is that the seed BH masses considered in our simulations are possibly on the higher end of the expected BH masses found in dwarfs, and hence our models might represent a small fraction of galaxies in this mass regime. 

Ultimately, distinguishing between these theoretical possibilities and constraining their respective frequencies requires a new frontier of observational work, extending dynamical measurements to lower masses to unveil whether there are dormant BHs in local dwarfs, and whether overmassive BHs are a widespread feature of low-mass galaxies in the early Universe or primarily arise from a combination of selection effects and current observational limitations.  


\subsubsection{$M_\bullet - \sigma_\star$ and $M_\bullet - M_\mathrm{dyn}$ relations}
\label{sec: M_BH vs M_dyn and sigma relations}

\begin{figure*}
\centering
\includegraphics[width=\linewidth]{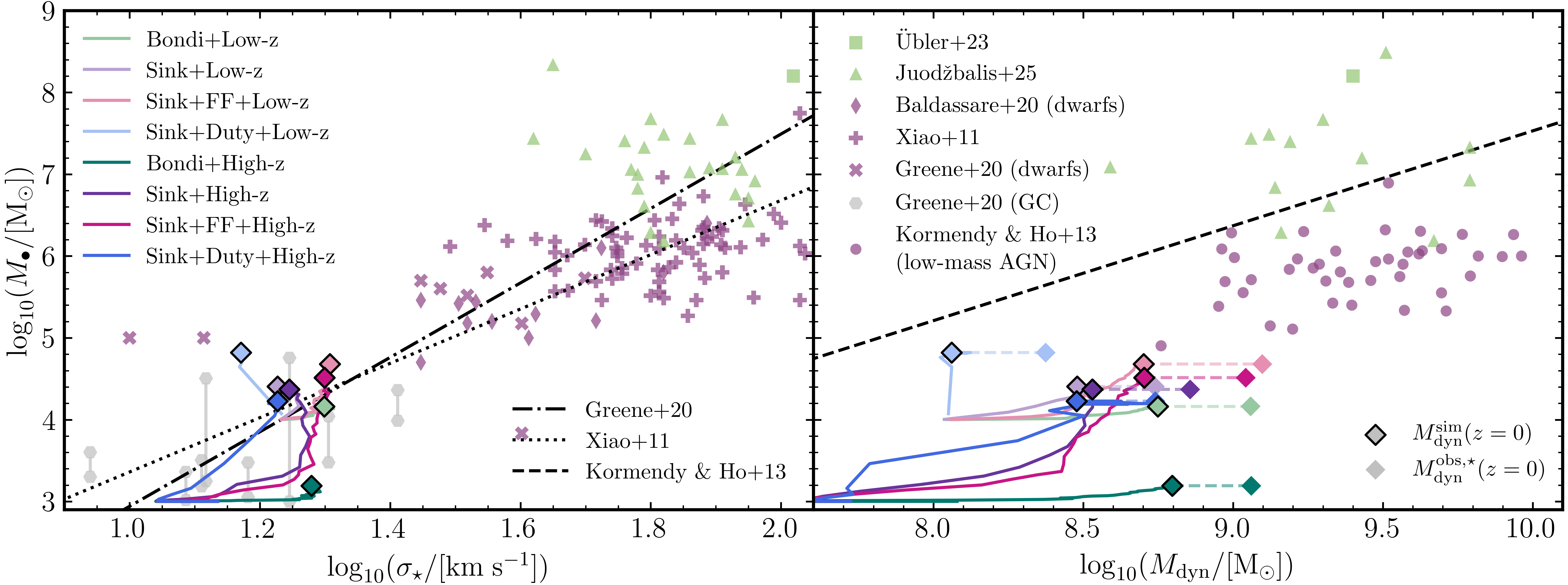}
\caption{\textit{Left-hand panel}: $M_\bullet - \sigma_\star$ scaling relation, for all simulation runs. \textit{Right-hand panel}: $M_\bullet - M_\mathrm{dyn}$ scaling relation, for all simulation runs. Simulation tracks are shown as solid lines. At each snapshot, the stellar effective radius $R_\mathrm{e}$ is derived from a Sérsic fit to the stellar surface mass density, and then used to compute $\sigma_\star$ and $M_\mathrm{dyn}$, as described in Section~\ref{sec: Mdyn estimators}. At $z=0$, the dashed line indicates the offset between the true dynamical mass ($M_\mathrm{dyn}^\mathrm{sim}$; black-edged diamonds) and the value computed via the \citet{van2022mass} estimator in Equation~(\ref{eq: vdW M_dyn}) with $\sigma_\star$ ($M_\mathrm{dyn}^\mathrm{obs,\star}$; diamonds without edges). For reference, we plot the extrapolated empirical scaling relations from \citet{xiao2011exploring} (black dotted line), \citet{greene2020intermediate} (black dash-dotted line) and \citet{kormendy2013coevolution} (black dashed line). We also report a sample of local and high-redshift data from low-mass galaxies (purple and green scatter points, respectively), as well as globular clusters (grey hexagons), as indicated in the legend. The simulated dwarfs generally follow the $M_\bullet - \sigma_\star$ relation due to moderate BH growth, but lie below the $M_\bullet - M_\mathrm{dyn}$ relation. This systematic bias between $M_\mathrm{dyn}^\mathrm{obs,\star}$ and $M_\mathrm{dyn}^\mathrm{sim}$ is likely due to gravitational potentials being heavily DM-dominated, while $R_\mathrm{e}$ traces only the stellar component. Despite significant differences in AGN feedback and star-formation histories, $\sigma_\star$ remains similar across runs, while variations in $R_\mathrm{e}$ primarily drive the scatter in $M_\mathrm{dyn}$.}
\label{fig: dynamical relations}
\end{figure*}

We now turn our analysis to the cosmic evolution of our simulation runs in $M_\bullet - \sigma_\star$ and $M_\bullet - M_\mathrm{dyn}$ phase space, as illustrated in the left- and right-hand panels in Figure~\ref{fig: dynamical relations}, respectively. For each run, the $z=0$ simulated data point is marked with a diamond. Similar to the $M_\bullet - M_\star$ analysis, we emphasise that the shape of these relations in the classical dwarf mass regime ($M_\star \sim 10^7 \; \mathrm{M_\odot}$) remains largely unconstrained due to observational limitations. Nevertheless, we extrapolate empirical scaling relations derived for higher-mass galaxies to investigate whether the trends observed in massive systems extend to our simulated dwarfs. Additionally, we plot available local and high-redshift observational data to place our simulations in the broader observational context.

Focusing on the $M_\bullet - \sigma_\star$ relation (left-hand panel), we show the fits from \citet{xiao2011exploring} (black dotted line) and \citet{greene2020intermediate} (black dash-dotted line). The \citet{xiao2011exploring} relation is based on a sample of Seyfert galaxies originally selected from the Sloan Digital Sky Survey and followed up with Keck spectroscopy, with $M_\bullet \gtrsim 2 \times 10^5 \; \mathrm{M_\odot}$, lying above the highest BH masses reached in any of our simulations at any redshift. Moreover, the \citet{greene2020intermediate} relation was derived using the dynamical BH masses from \citet{kormendy2013coevolution} and incorporating more recent dynamical BH mass measurements, including in the dwarf regime. Note that we plot the \citet{greene2020intermediate} relation based on the full dynamical sample. 

On the other hand, for the $M_\bullet - M_\mathrm{dyn}$ relation, we plot the \citet{kormendy2013coevolution} $M_\bullet - M_\star$ empirical scaling relation (black dashed line). This can be used as a proxy for the $M_\bullet - M_\mathrm{dyn}$ relation, as it is derived from a sample of early-type galaxies with low gas fractions, where the stellar mass accurately traces the dynamical mass \citep[also see][]{juodvzbalis2026jades}. 

Additionally, we plot observational data to show the scatter around the empirical relations, highlighting both the intrinsic dispersion and measurement uncertainties in the $M_\bullet - \sigma_\star$ and $M_\bullet - M_\mathrm{dyn}$ planes. Observed BH data points are colour-coded based on redshift, i.e. purple for low-redshift systems and green for high-redshift systems (i.e. \textit{JWST} data). For the low-redshift sample, $\sigma_\star$ measurements are derived directly from stellar absorption features, whereas in the high-redshift sample, $\sigma_\mathrm{gas}$ is used as a proxy for $\sigma_\star$, owing to observational limitations in obtaining reliable stellar kinematics from distant sources, as discussed in Section~\ref{sec: Mdyn estimators}. We include the sample of active dwarfs with virial BH masses from \citet{baldassare2020populating} (purple diamonds), where $\sigma_\star$ is measured from high-resolution Keck spectroscopy. Additionally, we plot the dwarfs from \citet{xiao2011exploring} (purple `+' symbols), as well as the dwarfs and globular clusters from  \citet{greene2020intermediate} (purple crosses and grey hexagons, respectively). As in Figure~\ref{fig: Mstar vs MBH}, we include error bars on $M_\bullet$ for the globular cluster sample, indicating the uncertainties in their BH mass measurements.
We also show the low-mass AGN sample from \citet{kormendy2013coevolution} as purple circles, which includes dynamical mass measurements. For the high-redshift regime, we include \textit{JWST} datasets from \citet{ubler2023ga} ($z \sim 5.55$, green square) and \citet{juodvzbalis2026jades} ($1.5 < z < 9.0$, green triangles), which were briefly described in Section~\ref{sec: M_BH vs M_star relation}. 

Focusing on the simulation tracks, at each snapshot we apply the Sérsic fitting procedure presented in Section~\ref{sec: Mdyn estimators} to derive the effective radius $R_\mathrm{e}$ and Sérsic index $n$, which we then use to compute $\sigma_\star$ and the true dynamical mass $M_\mathrm{dyn}$ (i.e. obtained by summing the masses of all particles and gas cells within $R_\mathrm{e}$ and multiplying by two). At $z=0$, the offset between the (median) true dynamical mass ($M_\mathrm{dyn}^\mathrm{sim}(z=0)$; black-edged diamonds) and the (median) value computed using the \citet{van2022mass} estimator in Equation~(\ref{eq: vdW M_dyn}) with $\sigma \equiv \sigma_\star$ (i.e. $M_\mathrm{dyn}^\mathrm{obs, \star}(z=0)$; diamonds with no edges) is indicated by a dashed line. 

The first thing to note is that our simulated dwarfs are largely in agreement with the $M_\bullet - \sigma_\star$ empirical relations, while they systematically lie below the $M_\bullet - M_\mathrm{dyn}$ relation. This behaviour can be explained by once again emphasising that our simulated dwarfs are heavily DM-dominated at all radii and throughout cosmic time. As a result, the value of $R_\mathrm{e}$ derived from the Sérsic fits does not correspond to the true half-mass radius of the entire galaxy, i.e. including the BH, gaseous, stellar and DM components. This leads to a fundamental discrepancy in the definition of $R_\mathrm{e}$, as our simulations and observational studies effectively measure $M_\mathrm{dyn}$ within different radii. 
This has a mild effect on the $M_\bullet - \sigma_\star$ relation, as $\sigma_\star$ is measured in a region where the velocity dispersion profile is relatively flat. In contrast, $M_\mathrm{dyn}$ has a much stronger dependence on the effective radius, leading to a mismatch between our simulated dwarfs and the empirical $M_\bullet - M_\mathrm{dyn}$ relation. Additionally, this effect is exacerbated for dynamical masses computed via Equation~(\ref{eq: vdW M_dyn}), as they scale linearly with $R_\mathrm{e}$ and with $\sigma_\star^2$ (itself mildly dependent on $R_\mathrm{e}$), while also relying on the prefactor $K(n)K(q)$, which is inaccurate due to the stellar component not tracing the overall gravitational potential in our dwarfs. 

Furthermore, despite our simulated dwarfs having significantly different star formation histories, BH accretion rates and gas content due to varying AGN feedback efficiencies (see Section~\ref{sec: cosmic evolution}), all simulation runs roughly cluster around similar values of $\sigma_\star$, with only moderate scatter reflecting their individual baryonic evolution. This effect is also visible in the $M_\bullet - M_\mathrm{dyn}$ relation, where the spread in $M_\mathrm{dyn}$ is larger because quenched systems (e.g. \textit{Sink+Duty+Low-z}) tend to have smaller $R_\mathrm{e}$, even though the underlying DM halo remains similar across different setups. 

Crucially, we find that there is no clear correlation between the characteristic growth histories of the BH, which are closely tied to the baryon cycle of the dwarf, and global dynamical properties such as $\sigma_\star$ and $M_\mathrm{dyn}$ in our simulations. Once again, this is a direct consequence of the gravitational potential being dominated by DM, which is largely insensitive to baryonic processes in our simulations. Specifically, all runs retain cuspy profiles at $z=0$, although we find a moderate trend towards lower central densities with increasing AGN feedback strength, consistent with \citet{koudmani2022two}. Nevertheless, our simulated dwarfs roughly follow the $M_\bullet - \sigma_\star$ relation, since BH growth is only moderate in most simulation runs, with the exception of the \textit{Sink+Duty+Low-z} run. 

Ultimately, it will be critical for future observational studies to probe the low-mass end of the $M_\bullet - \sigma_\star$ and $M_\bullet - M_\mathrm{dyn}$ scaling relations, where DM-dominated potentials likely play a key role. 


\subsection{Gas kinematics and metal enrichment}
\label{sec: outflows}

In this section, we aim to assess the extent to which AGN feedback can drive sustained outflows that ultimately have a large-scale impact on the baryon cycle of our simulated dwarfs. Specifically, we examine the kinematics and metal enrichment of the inflowing and outflowing gas for one representative setup, namely the \textit{Sink+Low-z} run. 

\begin{figure*}
    \centering
    \includegraphics[width=\textwidth]{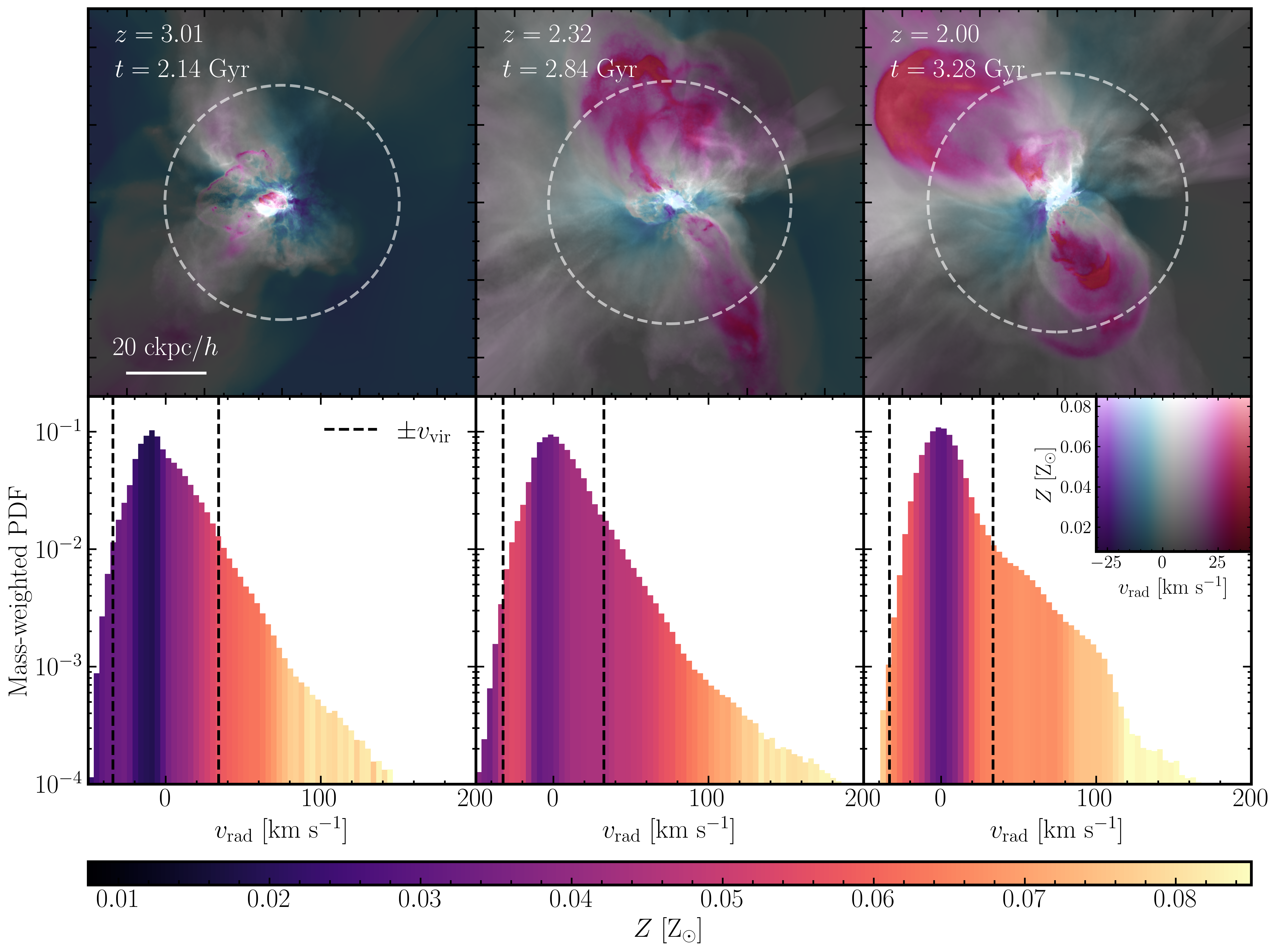}
    \caption{Kinematic and metal-enrichment properties of the gaseous component for the \textit{Sink+Low-z} run at three representative redshifts, i.e. $z = 3.01, 2.32$ and $2.00$.
    \textit{Upper panels:} Density-weighted, bivariate 2D projections of the gas metallicity and radial velocity. The projection dimensions are fixed at $(100 \; \mathrm{ckpc}/h)^3$. The region delimited by the virial radius is indicated by a white dashed circle. To simultaneously distinguish the inflowing and outflowing components and trace metal enrichment across cosmic time, we use a bivariate colourmap, which is shown in the upper right-hand corner of the third lower panel. The hue indicates the gas radial velocity and the brightness reflects the metallicity. \textit{Lower panels:} Mass-weighted PDF of the gas radial velocity, for the same snapshots as the upper panels. For reference, we plot $\pm v_\mathrm{vir}$ (where $v_\mathrm{vir}$ is the virial velocity) as vertical dashed black lines. Each bin is colour-coded based on its mass-weighted mean metallicity, as indicated by the linear colourbar. At early times, low-metallicity inflows feed the galaxy, while early AGN-driven outflows form confined, metal-enriched shells. As the system evolves, high-velocity, bipolar outflows extend beyond the virial radius, enriching both outflowing and inflowing gas through a galactic fountain effect, with the fastest outflows being the most metal enriched. }
    \label{fig: inflows vs outflows}
\end{figure*}

To this end, the upper panels in Figure~\ref{fig: inflows vs outflows} show bivariate, density-weighted projections of the total gas metallicity $Z$ and radial velocity $v_\mathrm{r}$ at three representative redshifts, namely $z=3.01$, $2.32$ and $2.00$, as indicated in the upper left-hand corner of each panel. This redshift range was chosen to capture the onset and subsequent evolution of large-scale, AGN-driven outflows in this run. In this work, we adopt the solar chemical composition values from \citet{asplund2009chemical}, for which $Z_\odot=0.0134$. Note that the latter is slightly higher than the fiducial value in \textsc{fable} and Illustris ($Z_\odot =0.0127$), although the exact choice of solar normalisation is subdominant compared to the systematic uncertainties inherent in nebular metallicity diagnostics, which can produce variations of more than a factor of two \citep[e.g.][]{kewley2008metallicity, torrey2019evolution, Maiolino_Mannucci2019}. In all panels, the projection dimensions are fixed at $(100 \; \mathrm{ckpc}/h)^3$, and the region delimited by the virial radius is indicated by a white dashed circle. To simultaneously distinguish inflows and outflows and trace metal enrichment across cosmic time, we adopt a bivariate colourmap, as shown in the upper right-hand corner of the third lower panel: the hue encodes the gas radial velocity, i.e. pink-red for outflows ($v_\mathrm{r}>0$) and blue-purple for inflows ($v_\mathrm{r}<0$), and the brightness reflects the gas metallicity, such that more metal-enriched regions appear brighter. Note that throughout this analysis, the radial velocity is computed as $v_\mathrm{r} \equiv \mathbf{v_\mathrm{cell}} \cdot 
\hat{\mathrm{r}}$, where $\mathbf{v_\mathrm{cell}}$ is the 3D velocity of each gas cell and $\hat{\mathrm{r}} \equiv \frac{(\textbf{r}_\mathrm{cell}-\textbf{r}_\mathrm{centre})}{|\textbf{r}_\mathrm{cell}-\textbf{r}_\mathrm{centre}|}$ is the radial unit vector. Bulk motions are removed by subtracting the mass-weighted mean gas velocity within twice the virial radius. 

To provide a more quantitative and complete picture of the kinematics and metal enrichment of inflows and outflows, the lower panels in Figure~\ref{fig: inflows vs outflows} show the mass-weighted probability density function (PDF) of the gas radial velocity at the same redshifts as the upper panels. Here, we only consider gas cells within twice the virial radius, and each bin represents the fraction of the total gas mass that sits within the corresponding velocity range. 
Each bin is colour-coded according to the mass-weighted mean metallicity of the gas cells falling within it. For reference, the virial velocity, i.e. $v_\mathrm{vir} = \sqrt{\mathrm{G} \frac{M_\mathrm{vir}} {R_\mathrm{vir}}}$, is marked with a vertical black dashed line.

At early cosmic times ($z=3.01$), most of the inflowing gas moves with low velocities ($|v_\mathrm{r}| \lesssim v_\mathrm{vir}$) and is metal-poor ($Z\sim0.01 \; \mathrm{Z_\odot}$), tracing the ongoing accretion of relatively pristine gas from the cosmic web. In the projections, these features are visible as dark-blue, low-brightness diffuse streams surrounding the central dwarf. In the innermost few kiloparsecs, most of the gas is metal-enriched and moves with low velocities (bright white patches), reflecting ongoing star formation activity in the central regions. Additionally, the first signatures of metal-enriched, AGN feedback-driven outflows (bright magenta patches) have already emerged from the central regions, forming metal-enriched shells that remain mostly confined within the virial radius. Note that these outflows can reach radial velocities of $v_\mathrm{r} \gg v_\mathrm{vir}$, indicating that some of the gas has sufficient kinetic energy to escape the host halo and enrich the CGM. The morphology changes dramatically by $z=2.32$, as large-scale, high-velocity bipolar outflows of metal-enriched gas (bright pink and red patches) extend well beyond the virial radius. Meanwhile, the central regions become increasingly filled with diffuse, moderately-enriched gas, while narrow, low-metallicity inflowing streams continue to supply the central regions. Furthermore, focusing on the velocity PDF at the same redshift, it is possible to identify a substantial metal-enriched inflowing component moving with $|v_\mathrm{r}| \lesssim v_\mathrm{vir}$, marking a characteristic galactic fountain effect, where previously ejected, metal-enriched gas `pollutes' the CGM and is then re-accreted as it falls back into the galaxy's gravitational potential well. This effect becomes more pronounced by $z \sim 2$, when powerful AGN-driven outflows reach their peak activity and suppress most of the inflowing material. At the same time, the gas becomes generally more metal-enriched across all velocities, reflecting the combined influence of AGN-driven outflows from the central regions and the galactic fountain cycle. As a result, both inflowing and outflowing gas are metal-enriched, with the latter component becoming increasingly dominant, as evident in the velocity PDF compared to earlier epochs.

Interestingly, we find that the fastest outflows are more metal-enriched than their slower-moving counterparts across all redshifts examined in Figure~\ref{fig: inflows vs outflows}. This trend naturally arises from our AGN feedback implementation, where energy is injected within a region capped at $R_\mathrm{sink}$ around the BH (see Section~\ref{sec: methods AGN feedback}). As a result, diffuse, metal-rich gas in the central star-forming regions is preferentially accelerated to high velocities. In contrast, slower outflows consist of denser clumps and gas located at larger radii, which are comparatively more metal-poor. Additionally, interactions during outflow propagation can further exacerbate this trend, as fast, metal-rich outflows mix with metal-poor CGM material, diluting their metallicity and reducing their velocity. Consequently, the slower component exhibits lower average metallicity, while the high-velocity tail retains the chemical signature of the central star-forming regions. We note, however, that our simulations do not include radiation pressure on dust, which would preferentially accelerate metal-rich (and hence dust-rich) gas, potentially contributing to the observed metallicity-velocity correlation. 

An additional feature to note in the velocity PDFs is the dark, low-metallicity peak around $v_\mathrm{r} \sim 0$. This is visible across all redshifts and arises from our mass-weighting procedure, which includes all gas within $2 \times R_\mathrm{vir}$, such that the contribution from diffuse, metal-poor CGM gas at large radii dominates the very low-velocity regime. In particular, this component originates predominantly from directions perpendicular to the main outflow axes, where the gas is less exposed to AGN feedback (low-brightness, grey patches in the projections) and thus retains its low metallicity, although the inner regions evolve considerably across cosmic time. However, note that the peak becomes progressively lighter towards lower redshifts, indicating a gradual enrichment of the diffuse CGM as outflows and recycled material mix with the surrounding halo gas.

As a final remark, we emphasise that metal enrichment of the outer CGM, i.e. around and beyond the virial radius, is a feature of all efficient AGN runs, with outflows being launched at a redshift that broadly coincides with the onset of star formation quenching, thus implying that the epoch of CGM metal enrichment depends sensitively on the timing of peak AGN activity. This behaviour is further explored in Appendix~\ref{sec: appendix B}, which shows that early, efficient AGN activity (i.e. in the \textit{Sink+Duty+High-z} run) can produce substantial metal enrichment in the outer CGM, which persists down to $z=0$, as opposed to inefficient AGN runs (i.e. the \textit{Bondi+High-z} run). Such long-lived CGM metal enrichment could therefore serve as a fossil record of past BH growth and AGN activity, even in systems where the BH is currently dormant and star formation has been quenched. However, owing to the extremely low column densities involved in this dwarf system, this material would be challenging to detect with current observational facilities, although it may be more readily observable in higher mass galaxies. 


\section{Discussion}
\label{sec: discussion}


\subsection{Comparison with previous theoretical work}
\label{sec: comparison with previous work}

The results obtained in this work show that, provided that the BH is allowed to accrete efficiently, i.e. via an accretion prescription that does not by definition penalise the growth of low-mass BHs, AGN feedback can play a crucial role in shaping the baryon cycle of its dwarf host. This behaviour differs significantly from a large body of cosmological simulations, typically employing Bondi-based BH accretion models in combination with strong SN feedback parametrisations, which find that BH growth is effectively suppressed in the dwarf regime \citep[e.g.][]{dubois2015black, habouzit2017blossoms, trebitsch2018escape}.

Instead, our work is in broad agreement with recently emerging studies which consider more efficient accretion pathways onto low-mass BHs. In particular, this work is most directly comparable to the one from \citet{koudmani2022two}, as both studies are based on the \textsc{fable} galaxy formation model and perform zoom-in simulations of the same dwarf galaxy. Similar to \citet{koudmani2022two}, exploring BH accretion models beyond Bondi-based schemes allows us to enter the regime of early ($z \gtrsim 2$), efficient low-mass BH growth, although here this is achieved via a novel sink-based model rather than the supply-limited approach adopted in their work, which initiates Eddington-limited bursts whenever the gas supply is sufficient. Compared to \citet{koudmani2022two} we implement high-redshift BH seeding ($z_\mathrm{seed} \sim 15$) in half of our simulation runs, such that our models interface more naturally with the early-Universe BH population now being unveiled by \textit{JWST}.

Furthermore, our results broadly align with trends seen in a number of other cosmological simulations of dwarf galaxies. For example, \citet{sharma2020black} analyse a sample of 205 isolated dwarfs in the \textsc{romulus25} simulation and find that dwarfs that host overmassive BHs relative to the mean $M_\bullet - M_\star$ relation display early ($z\sim 2$) suppression of star formation as a result of efficient AGN activity. Similarly, \citet{koudmani2021little} analyse the \textsc{fable} simulation suite and find that overmassive BHs can drive hot, accelerated outflows that considerably drain the gas reservoir of the dwarf, thus playing a role in the quenching of dwarfs at high redshift. The same work shows that the fiducial \textsc{fable} setup might underestimate the number of X-ray bright AGN in dwarfs due to the strong SN feedback implementation. 

On the other hand, there is also a body of theoretical work that suggests that the BH accretion rates in dwarfs are generally low, especially if the BH is off-centre and wanders in the shallow potential well of its host \citep[e.g.][]{ bellovary2019multimessenger, pfister2019erratic}. \citet{sharma2022hidden} find that there is a hidden population of massive BHs (many of which are undermassive compared to their hosts and off-centre) with low X-ray luminosities in dwarf galaxies, which are below the current X-ray detection limits ($L_\mathrm{X}^\mathrm{AGN}< 10^{39}$ erg s$^{-1}$) and suffer from contamination from XRBs, among other sources. 

This variety of outcomes is largely imputable to the plethora of subgrid models used to account for BH physics in cosmological simulations, which are all particularly sensitive to the choice of the BH seed mass, specific BH accretion and AGN feedback prescriptions, as well as tracking of BH dynamics. In general, simulations still struggle to produce results in agreement with observational constraints, such as the observed occupation fraction of X-ray bright AGN in dwarfs and the stellar mass--halo mass relation. This challenge is further exacerbated by uncertainties in observational measurements and the scarcity of direct observational constraints in this galaxy mass domain.  

\subsection{Comparison with local dwarf AGN}

The local population of dwarf AGN provides an important observational context for interpreting our results, although the stellar mass regime probed in this work ($M_\star \lesssim 10^{8} \, \mathrm{M_\odot}$) remains sparsely sampled and subject to selection effects, such that the demographics of dwarf AGN remain poorly understood. Nevertheless, the census of dwarf AGN candidates has substantially expanded through complementary multi-wavelength diagnostics \citep[e.g.][]{reines2013optical, kaviraj2019agn, reines2020new, birchall2020x, molina2021sample, polimera2022resolve, bykov2024erosita, sacchi2024x, Pucha2025, pucha2026desi} and variability searches \citep[e.g.][]{baldassare2020variability, burke2022variability, kaviraj2025variability}, with recent DESI data more than tripling the existing census \citep[][]{Pucha2025, pucha2026desi}. The inferred active fraction, however, remains diagnostic- and sample-dependent and provides only a lower limit on the BH occupation fraction. In particular, standard optical diagnostics can miss AGN in low-metallicity, highly star-forming dwarfs \citep{polimera2022resolve}, while shallow surveys preferentially select brighter, actively star-forming systems \citep{kaviraj2025quenching}.

Although current samples are dominated by blue, star-forming galaxies, \citet{lazar2026downsizing} find that downsizing trends do not trivially extend to the dwarf regime, with the red fraction rising below $M_\star \sim 10^{8.5} \, \mathrm{M_\odot}$, including in low-density environments, pointing towards internally driven quenching \citep[see also][]{kaviraj2025quenching, bidaran2025quenching}. While they identify SN feedback as the primary driver of quenching, detections of fast, spatially extended outflows associated with dwarf AGN indicate that AGN feedback may also influence their gas content \citep[e.g.][]{manzano2019agn, liu2020integral, zheng2023escaping, morales2025manga}.

Despite these advances, observational and modelling uncertainties, together with the zoom-in, single-halo nature of our simulation suite, make direct comparisons particularly challenging and necessarily qualitative. We emphasise that our aim is not to reproduce the local dwarf AGN demographics or validate a particular model, but rather to isolate the impact of different subgrid BH accretion prescriptions on the evolution of the host. The very low final stellar masses ($M_\star \sim 10^6 \, \mathrm{M_\odot}$) in the \textit{Sink+Duty} setups should therefore be interpreted as bracketing the strong-feedback extreme rather than representing a typical evolutionary pathway for dwarf AGN hosts, consistent with the similarly catastrophic outcomes found in \citet{koudmani2022two}. Although these runs fall below extrapolated $M_\star - M_\mathrm{halo}$ relations, it should be emphasised that these remain poorly constrained at such low stellar masses \citep[e.g.][]{garrison2017smhm, munshi2021smhm}.

At the other extreme, the \textit{Bondi} and \textit{Sink+FF} setups remain star forming down to $z=0$, qualitatively resembling the star-forming systems prevalent in current dwarf AGN samples. Combined with the limitations discussed above, these contrasting outcomes demonstrate that the impact of AGN feedback does not map onto a simple Bondi-versus-sink distinction, preventing current dwarf AGN demographics from discriminating between or definitively validating either prescription. Determining whether low-mass, quenched dwarfs containing dormant, overmassive BHs exist in the local Universe, and how frequently they occur, requires observations below the stellar-mass range of present dwarf AGN samples, as well as statistically representative cosmological simulations, both of which lie beyond the scope of the present study. 

The structural properties of our simulated dwarfs provide an additional point of qualitative comparison with morphological studies of the local dwarf galaxy population. Despite markedly different evolutionary histories, most runs retain globally near-exponential surface density profiles (median Sérsic index $n \sim 0.8-1.4$), with the \textit{Sink+Duty+High-z} showing the clearest departure from this behaviour (see Appendix~\ref{sec: appendix A} for further details). This is qualitatively consistent with the predominantly exponential profiles measured in a large sample of low-mass dwarfs by \citet{kadofong2020morphology}, although their data is not AGN-selected. We therefore find no robust separation in global, single-component morphology among the BH accretion prescriptions, although this conclusion is necessarily limited by our simplified morphological analysis. Interestingly, recent morphological studies suggest that the central regions of dwarfs may provide a more informative diagnostic. \citet{kimbrell2023morphology} find that AGN dwarfs with $10^9 \, \mathrm{M}_\odot \lesssim M_\star \lesssim 10^{9.5} \, \mathrm{M}_\odot$ commonly present a central (pseudo)bulge component absent in many inactive systems, with similar structures reported in variability-selected hosts at lower stellar masses \citep{kimbro2025morphology}. However, our eEOS treatment of the ISM, the absence of NSC modelling, and our global, single-component Sérsic fits preclude an equivalent decomposition. We therefore identify this as an important direction for future simulation-based work, further motivating the theoretical advances discussed in Section~\ref{sec: caveats}. 

Ultimately, future observations will be essential for determining where real dwarf
galaxies lie within the range of behaviours identified in this work. Improved constraints on the low-mass AGN occupation fraction, together with deeper surveys of isolated quenched dwarfs and measurements of circumgalactic metal enrichment, will provide powerful tests of BH growth and feedback models in the dwarf galaxy regime. In addition, gravitational-wave observations with future facilities such as \textit{LISA} will place independent constraints on the occupation and merger histories of IMBHs.


\subsection{Caveats and future work}
\label{sec: caveats}

As in any numerical study, the simulations analysed in this work are subject to caveats associated with resolution limits and modelling assumptions. In particular, BH accretion and the subsequent coupling of AGN feedback to the host galaxy operate across a vast range of spatial and temporal scales, thus necessitating that several key physical processes be treated in a subgrid fashion, rather than modelled from first principles. In this section, we highlight the main limitations of this approach to help contextualise and interpret our results.


\subsubsection{Black hole seeding, dynamics and accretion}
\label{sec: BH modelling caveats}

In this work, we leverage high-resolution cosmological simulations of an individual dwarf galaxy, enabling us to accurately track gas inflows onto the central BH and systematically study the sensitivity of the system to variations in our novel subgrid sink-particle BH accretion scheme (see Section~\ref{sec: methods sink-particle model}). While this approach enables highly controlled numerical experiments, the computational cost restricts us to a narrow range of BH seeding scenarios and other model parameters, preventing exploration of the broader diversity of BH populations and environments that would be accessible in large-volume cosmological simulations. Since our aim is to investigate the possibility of efficient BH accretion in dwarfs, we intentionally focus on BH seed masses at the high-mass end of what may be expected for galaxies of comparable stellar mass (see Figure~\ref{fig: Mstar vs MBH}), although observational constraints in this regime remain extremely scarce. As a consequence, BH growth is potentially biased towards relatively efficient early accretion, which may not be representative of the full population of BHs in dwarf galaxies. 

Furthermore, while our simulated dwarf resides in a low-density, relatively quiet environment, it is crucial that future work accounts for the diverse range of possible cosmological environments and merger histories of dwarf galaxies, alongside improved modelling of BH dynamics. Indeed, both theoretical and observational studies find that perturbations such as galaxy mergers may lead to off-centre BHs, which might spend the majority of their orbits in low-density regions and never sink to the nuclear region of the dwarf \citep[e.g.][]{reines2020new, bellovary2021origins}. These BHs undergo significantly fewer accretion events compared to their nuclear counterparts, resulting in negligible impact on the star formation history of their hosts. This calls for a more realistic treatment of BH orbital decay towards the galactic centre, which is dictated by dynamical friction from the surrounding gas, stars and DM. Accurately capturing this process will be crucial to enhance the predictive power of cosmological simulations, and would likely help shed light on the multitude of results produced by current numerical studies, in which AGN activity seems to be mostly restricted to high redshifts ($z>2$), whilst local dwarfs are mostly characterised by a population of dormant BHs \citep[e.g.][]{sharma2022hidden, koudmani2021little}. 

In addition to dynamical friction modelling, future work will need to focus on more sophisticated models for BH orbital decay and spin evolution in response to feedback-regulated gas flows from the multi-phase ISM \citep[e.g.][]{dubois2014black, fiacconi2018galactic, cenci2021black, bollati2024exploring, koudmani2024unified}. Although this remains a major challenge due to the vast range of dynamical scales involved \citep[e.g.][]{shin2025mandelzoom}, incorporating post-Newtonian solvers such as \textsc{ketju} \citep[][]{rantala2017post, mannerkoski2023ketju} into high-resolution hydrodynamical galaxy simulations offers a promising path forward. 

Another fundamental caveat of this work concerns our novel treatment of BH accretion, which is implemented via a simple subgrid sink-particle scheme (see Section~\ref{sec: methods sink-particle model}). We emphasise once again that the numerical experiments presented here are primarily aimed at exploring whether alternatives to Bondi-based models can lead to efficient BH growth and AGN feedback in simulated dwarfs, rather than to provide a definitive description of BH accretion applicable across the entire galaxy mass range. 

Furthermore, note that the expression for the free-fall timescale used in these simulations (see Equation~(\ref{eq: ff time})) only accounts for the mass of the BH. A more accurate treatment would include the  total mass of gas, stars and DM within the sink radius, which would naturally result in a lower value of $\tau_\mathrm{FF}$ compared to only considering the BH mass, as $\tau_\mathrm{FF} \propto M^{-1/2}_{r < R_\mathrm{sink}}$. Nevertheless, the dependence of the free-fall timescale on the enclosed mass is weaker than its dependence on the sink radius ($\tau_\mathrm{FF} \propto R_\mathrm{sink}^{3/2}$), such that implementing this refinement would likely still attenuate differences arising from varying $R_\mathrm{sink}$.

While our sink-particle method is computationally efficient and does not artificially suppress the growth of low-mass BHs, it neglects several important physical factors, including the angular momentum of the infalling gas and the detailed multi-phase structure of the ISM. This calls for the exploration of different subgrid BH accretion models in simulations of dwarf galaxies, including gas-supply- and torque-limited schemes \citep[e.g.][]{angles2013black, angles2015torque, koudmani2022two, wellons2023exploring, gordon2024hungry}, as well as disc-based approaches \citep[e.g.][]{power2011accretion, fiacconi2018galactic, dubois2014black, cenci2021black, koudmani2024unified}. In particular, the latter models are able to resolve small-scale mass and angular momentum flows onto a subgrid accretion disc, allowing for the self-consistent evolution of the BH spins, but require resolving much smaller spatial scales, hence significantly increasing the computational cost. 

In particular, such disc-based approaches have recently been applied to idealised simulations of dwarf galaxies in \citet{shin2025mandelzoom, shin2026mandelzoom2}, resolving mass and angular momentum transport onto a subgrid $\alpha$-disc down to $\lesssim 0.01 \, \mathrm{pc}$ scales. Although the idealised, non-cosmological nature of these setups precludes a direct comparison with our own, they demonstrate that BH growth in dwarfs is strongly sensitive to the treatment of angular momentum transport in the nuclear region, as well as to the stellar feedback channels regulating the circumnuclear gas supply, which cannot be resolved in cosmological zoom-in simulations at the resolution adopted here. Ultimately, we regard the two approaches as complementary: high-resolution, idealised studies resolve the nuclear-scale physics setting the gas supply to the BH, whereas the sink-based prescriptions adopted here are designed to quantify the uncertainty associated with unresolved accretion physics within a fully cosmological setting. Incorporating physically motivated constraints from the former therefore represents an important direction for future developments of subgrid accretion models in cosmological simulations. 

In addition to improved subgrid models, recent developments in state-of-the-art multi-scale computational techniques provide a complementary route to address these modelling limitations. These include zoom-in nested simulations \citep[e.g.][]{hopkins2010massive, kaaz2025h}, super-Lagrangian refinement techniques \citep[e.g.][]{curtis2015resolving, angles2021cosmological}, extended general relativistic magnetohydrodynamic (GRMHD) simulations \citep[e.g.][]{lalakos2022bridging, galishnikova2025strongly} and `multi-zone' computational methods \citep[e.g.][]{cho2023bridging, cho2024multizone}. Although these approaches are beginning to bridge the vast dynamic range between galaxy-scale and GRMHD simulations, a fully self-consistent treatment across all relevant scales has yet to be achieved. 

Furthermore, several theoretical studies have pointed out the importance of tidal disruption events in driving the growth of IMBHs in dense stellar environments, such as NSCs \citep[e.g.][]{stone2017formation, zubovas2019tidal, lee2023growth, rizzuto2023growth, rantala2024frost, chang2025rates}. Additionally, the inclusion of a NSC can significantly aid the funnelling of gas towards the central BH in idealised simulations of dwarf galaxies \citep[e.g.][]{partmann2025importance, shin2025mandelzoom}. In future work, it will therefore be crucial to include this BH growth channel by modelling realistic NSC density profiles and stellar dynamical processes in cosmological simulations. 


\subsubsection{Baryonic feedback}
\label{sec: baryonic feedback caveats}

An additional caveat concerns the treatment of AGN feedback, which here is modelled exclusively in the form of isotropic thermal injections in the central region surrounding the BH. This simplification is deliberate, as this work focuses on isolating the effects of varying the BH accretion model, rather than providing a comprehensive study of AGN feedback. Nevertheless, while the thermal energy injection may mimic the effect of energy-conserving outflows expected from fast AGN winds \citep[e.g.,][]{zubovas2012clearingout, bourne2013energyconserving, costa2014feedback}, more realistic AGN feedback prescriptions that leverage super-Lagrangian refinement techniques could be implemented to account for different injection geometries, which could potentially enhance BH growth by preserving the gas reservoir in the central regions. In particular, including processes that are specific to disc-like structures, such as disc winds \citep[e.g.][]{zubovas2016bipolar, koudmani2019fast, sala2021non} and collimated jets \citep[e.g.][]{bourne2017agn, bourne2019agn, mukherjee2018jetdisc, talbot2021blandford, su2021agn} may significantly affect the interplay between AGN feedback and the ISM in simulated dwarfs.

Beyond AGN feedback, an important caveat is that we have explored only one parametrisation of SN feedback. Stronger parametrisations might remove gas from the vicinity of the BH, preventing efficient BH accretion \citep[see e.g. ][for a more detailed discussion]{koudmani2022two}. Future work should strive to improve the modelling of stellar feedback in cosmological simulations of dwarfs, focusing on approaches that prioritise explicit IMF sampling schemes to capture the stochastic, bursty nature of SN explosions instead of distributing feedback continuously over time \citep[e.g.][]{applebaum2020stochastically, smith2021efficient, shin2026mandelzoom2}. 

A further limitation of the present work concerns the treatment of radiative feedback. In our simulations, ionising radiation is included only in the form of a spatially uniform, redshift-dependent UV background, as inherited from Illustris and \textsc{fable} \citep{vogelsberger2013model, henden2018fable}, with no local radiative transfer of stellar or AGN photons. The need for ISM physics and other feedback processes beyond SN feedback alone has been previously highlighted in \citet{smith2019cosmological} using the same initial conditions adopted here, and dwarf-galaxy predictions are known to be highly sensitive to this modelling in a non-trivial way, with local stellar radiation affecting both star formation histories and SN-driven outflows \citep[e.g.][]{baumschlager2025, rey2025edge}. The net effect of radiative feedback on BH fuelling and subsequent AGN feedback is therefore not obvious. Recent idealised simulations of dwarf galaxies by \citet{shin2026mandelzoom2} suggest that local stellar radiation can enhance BH fuelling by suppressing fragmentation and inducing the inward compression of circumnuclear gas, although its coupling to SN feedback makes accretion highly episodic. Similarly, \citet{petersson2025noctua} find that ionising stellar radiation enhances BH growth by suppressing star formation and increasing the net gas inflow, while partially offsetting the suppression caused by SNe. Disentangling these effects in a fully cosmological context, along with their interplay with AGN feedback itself, would require a resolved multiphase ISM alongside a model for local ionising radiation, and represents an important direction for future work.

Additionally, an increasing body of theoretical work on dwarf galaxies underscores the importance of accurately modelling the interaction between AGN feedback and non-thermal components such as CRs, which may considerably aid AGN-driven suppression of star formation \citep[e.g.][]{su2021agn, wellons2023exploring, martin2023pandora, su2024unravelling, byrne2024effects, koudmani2025diverse, martin2026pandora}. 

Accurately capturing this interplay will therefore be critical for understanding the diverse pathways through which baryonic feedback processes jointly regulate the cosmic evolution of dwarf galaxies.  


\subsubsection{ISM physics}
\label{sec: ISM physics caveats}

Finally, the unresolved ISM is a major caveat for the simulations explored in this work. Here, the multi-phase structure of the ISM is approximated via an eEOS (see Section~\ref{sec: methods SF and ISM}), following the approach originally presented in \citet{springel2003cosmological}. Modelling the cold phase of the ISM will be essential to accurately assess the efficiency of BH accretion and the subsequent impact of AGN in dwarfs, as dense, cold clumps could potentially contribute to feeding the BH. Additionally, the energy from AGN feedback would likely couple differently with a multi-phase ISM relative to the subgrid modelling described by the eEOS \citep[see e.g.,][]{wagner2012jetsmultiphase, wagner2013ufomultiphase, bourne2014multiphase, bourne2015resolution, mukherjee2018jetdisc}. In these regards, it is important to note that the effects of a resolved ISM model on BH accretion rates and AGN feedback efficiencies are not readily evident, calling for the need for future work. On the one hand, some studies find that multi-phase ISM models might result in considerably more spatially localised and bursty star formation and stellar feedback activity, which could hinder BH growth by expelling gas from the host \citep[e.g.][]{angles2017black}. On the other hand, AGN feedback could interact with a resolved ISM by expelling hot gas faster than cold clumps, with the latter fuelling BH accretion \citep[e.g.][]{angles2021cosmological, sivasankaran2022simulations}. Furthermore, although the implementation of a resolved ISM model would be a fundamental step forward, AGN- and SN-driven outflows should still be clearly distinguishable, with AGN-driven outflows displaying higher velocities and temperatures \citep[e.g.][]{koudmani2019fast}. Lastly, the effects of a clumpy, resolved ISM on the dynamics of the BH make improved models of dynamical friction even more critical.


\section{\textsc{Conclusions and future outlook}}
\label{sec: conclusions}

Dwarf galaxies are ideal laboratories to study the early seeding of BHs and the co-evolutionary growth with their host galaxies. Observationally, \textit{JWST} is uncovering a population of low-mass BHs in high-redshift galaxies, while surveys such as MaNGA and VIPERS are probing AGN activity in dwarfs at low-to-intermediate redshifts. Theoretically, many cosmological simulations of galaxy formation might penalise the growth of low-mass BHs and underestimate the impact of AGN in simulated dwarfs, owing to a combination of strong SN feedback parametrisations and simplistic Bondi-based BH accretion models. 

Motivated by this, we perform high-resolution cosmological zoom-in simulations of an individual dwarf galaxy based on \textsc{fable} physics to explore whether moving beyond Bondi-based models can enable efficient accretion onto low-mass BHs and enhance AGN feedback. We introduce a novel sink-based BH accretion model, in which BH accretion is treated by means of an effective collisionless \textit{sink particle}, with the accretion rate evaluated directly from the resolved gas distribution within a finite spatial region surrounding the BH. We then consider modifications of this basic scheme, including a duty cycle to capture episodic accretion, as well as a free-fall limited prescription. For each model, we explore two BH seeding scenarios: a \textit{Low-z} setup ($z_\mathrm{seed} \sim 4$ and $M_\mathrm{seed} \sim 10^4 \, \mathrm{M}_\odot$) and a \textit{High-z} setup ($z_\mathrm{seed} \sim 15$ and $M_\mathrm{seed} \sim 10^3 \, \mathrm{M}_\odot$), which allow us to focus on the regime of efficient BH accretion and AGN feedback, as these seed masses likely lie at the more massive end of BHs typically found in dwarfs.

We analyse the simulations across a broad range of diagnostics, including the cosmic evolution of properties related to BH growth, galaxy assembly and star formation, the location of our simulated dwarfs on well-established scaling relations, and the characteristics of AGN-driven outflows and their effect on the ISM and CGM. Our main findings are summarised below:

\begin{enumerate}
        \item There is enough gas in the nuclear regions of the dwarf to fuel efficient BH accretion, as seen in all sink-based runs. Nevertheless, Bondi-based models yield low BH accretion rates that result in negligible impact of AGN feedback on the baryon cycle of the dwarf. This is in agreement with the findings of \citet{koudmani2022two}, and it suggests that this class of models is inadequate for use in cosmological simulations of AGN in dwarf galaxies;
        \item The impact of AGN feedback on the star formation activity of the dwarf is dependent on the specific BH accretion implementation, ranging from negligible impact (\textit{inefficient} AGN runs) to dramatic quenching at high redshift ($z\gtrsim 2$; \textit{efficient} AGN runs). The inclusion of a $25 \; \mathrm{Myr}$ AGN duty cycle leads to extremely efficient BH accretion at early times by delaying the onset of AGN feedback. This allows more efficient coupling between AGN feedback and the ISM compared to the other setups. Furthermore, introducing the free-fall timescale as a limit to the BH accretion rate results in more moderate and sustained BH accretion across cosmic time, with no global quenching. These effects are observed regardless of the specific BH seeding prescription adopted; 
        \item We tested the performance of two virial estimators for the dynamical mass, namely the \citet{walker2009universal} and the \citet{van2022mass} estimators,
        against the true dynamical mass of our simulated dwarfs. At low redshifts, when the stellar velocity dispersion is used, our results indicate that while the \citet{walker2009universal} estimator accurately recovers the true dynamical mass of the galaxy to within $\sim 10-15 \%$, the \citet{van2022mass} estimator, commonly employed in \textit{JWST}-based analysis, systematically overestimates the true dynamical mass by a factor of $\sim 2$. Both estimators struggle to reproduce the true dynamical mass of the galaxy at high redshift, as the system is not dynamically relaxed yet, which may be the case for a number of \textit{JWST}-observed early systems. Furthermore, as the gas kinematics is very sensitive to AGN feedback, the gas velocity dispersions should be used with caution to estimate the dynamical mass;
        \item Focusing on the evolution of the simulated dwarfs in $M_\bullet-M_\star$ phase space, at $z=0$ efficient AGN runs produce BHs that, for a given stellar mass, are overmassive relative to the extrapolated local scaling relation from \citet{reines2015relations}, while falling within the scatter of the high-redshift relation from \citet{pacucci2023jwst}. In contrast, runs with inefficient AGN feedback yield BHs that remain consistent with the local scaling relation at $z=0$. 
        Thus, our simulations bracket two potential evolutionary pathways: one leading to an elusive local population of dormant, overmassive BHs as remnants of efficient BH accretion episodes in high-redshift galaxies (efficient AGN runs), and another where stellar build-up catches up with BH growth as cosmic time progresses (inefficient AGN runs);
        \item As our simulated dwarfs are DM-dominated at all times and radii, there is no clear correlation between the growth histories of the central BHs and global dynamical properties of the galaxy, including the stellar velocity dispersion and the dynamical mass. In fact, while the latter are primarily determined by the DM potential, BH growth and AGN self-regulation are closely tied to the baryon cycle of the dwarf, which has very little influence on the DM component in our simulations;  
        \item Runs with efficient AGN feedback drive metal-enriched outflows from the central regions of the dwarf, which extend well beyond the virial radius. These outflows deplete the central gas reservoir of the dwarf and pollute the CGM. Infrequent AGN feedback bursts are sufficient to maintain hot gas at the scale of $R_\mathrm{vir}$, suppress cosmic inflows and keep the dwarf in a quenched state;
        \item In setups with efficient AGN feedback, metal enrichment in the outer CGM, i.e. around and beyond the virial radius, is primarily driven by AGN outflows and may thus provide a fossil signature of past BH growth and AGN feedback activity, even in systems with dormant BHs and quenched star formation activity. However, this material is extremely challenging to constrain with current observational facilities.  
\end{enumerate}

In conclusion, this work demonstrates that moving beyond simplistic Bondi-based BH accretion prescriptions, in this case via sink-based models, unlocks the largely unexplored regime of efficient BH accretion and powerful AGN feedback in simulated dwarf galaxies. While this scenario has received little theoretical investigation until recently, it is now increasingly motivated by mounting observational evidence at all redshifts. By systematically comparing different, simple subgrid prescriptions for BH accretion, we quantify how alternative treatments of unresolved gas inflow lead to significantly different BH growth and feedback histories in dwarf galaxies, despite an otherwise consistent simulation framework.
This work thus represents a step forward in explaining these observations, although it does not provide a definitive model for BH accretion in dwarfs. It is therefore critical that the next generation of cosmological simulations of dwarf galaxies incorporates more realistic and self-consistent models for BH accretion, BH dynamics, ISM physics and baryonic feedback. Equally important will be the development of novel multi-scale computational strategies to efficiently explore the high-dimensional parameter space and improve the physical fidelity and scalability of future simulations. Alongside these theoretical and methodological advances, future observational efforts should be directed towards extending dynamical BH mass measurements to lower masses and higher redshifts. Together, such approaches will help determine whether overmassive BHs in local dwarf galaxies and those probed by \textit{JWST} at high redshift represent manifestations of a common, potentially widespread mode of rapid BH growth in low-mass systems at early times, therefore providing a unique window into the origin and evolution of BHs in the early Universe.

\section*{Acknowledgements}
The authors are grateful to Matthew C. Smith for granting access to the \textsc{vortrace} module for producing the projections. G.O. acknowledges support from a Gates Cambridge Scholarship, a Cambridge Scholarship for Italian Talent, an Ermenegildo Zegna Founder's Scholarship, and an Institute of Astronomy summer internship bursary.
M.A.B. is supported by a UKRI Stephen Hawking Fellowship (EP/X04257X/1). D.S. acknowledges support from the Science and Technology Facilities Council (STFC) under grant ST/W000997/1.

\section*{Data Availability}
 
The data underlying this article will be shared on reasonable request to the corresponding author.


\bibliographystyle{mnras}
\bibliography{refs} 



\appendix

\section{Further diagnostics on the validation of dynamical mass estimators}
\label{sec: appendix A}

In Section~\ref{sec: Mdyn estimators}, we analysed the evolution of the dynamical mass across cosmic time, based on different virial estimators, focusing on the \textit{Sink+Low-z} run. Specifically, we compared the true value of the dynamical mass, i.e. $M_\mathrm{dyn}^\mathrm{sim}$, with estimates using Equation~(\ref{eq: vdW M_dyn}) (\citet{van2022mass} estimator) and Equation~(\ref{eq: Walker M_dyn}) (\citet{walker2009universal} estimator),  i.e. $M_\mathrm{dyn}^\mathrm{obs,\star}$ (setting $\sigma \equiv \sigma_\star$) and $M_\mathrm{dyn}^\mathrm{obs,gas}$ (setting $\sigma \equiv \sigma_\mathrm{gas}$). In Figure~\ref{fig: M_dyn vs t}, we noted how methods based on Equation~(\ref{eq: vdW M_dyn}) systematically overestimate the true dynamical mass at all times, regardless of the efficiency of AGN feedback. One possible explanation for this discrepancy is that our simulated dwarf galaxies are DM-dominated at all times and radii. As a consequence, their stellar surface density profiles may not trace the underlying gravitational potential within the effective radius $R_\mathrm{e}$ in the same way as the galaxies used to calibrate the \citet{van2022mass} dynamical mass estimator.


\subsection{Radial density profiles for the \textit{Sink+Low-z} run}
\label{sec: radial density profiles appendix}

\begin{figure*}
    \centering
    \includegraphics[width=\textwidth]{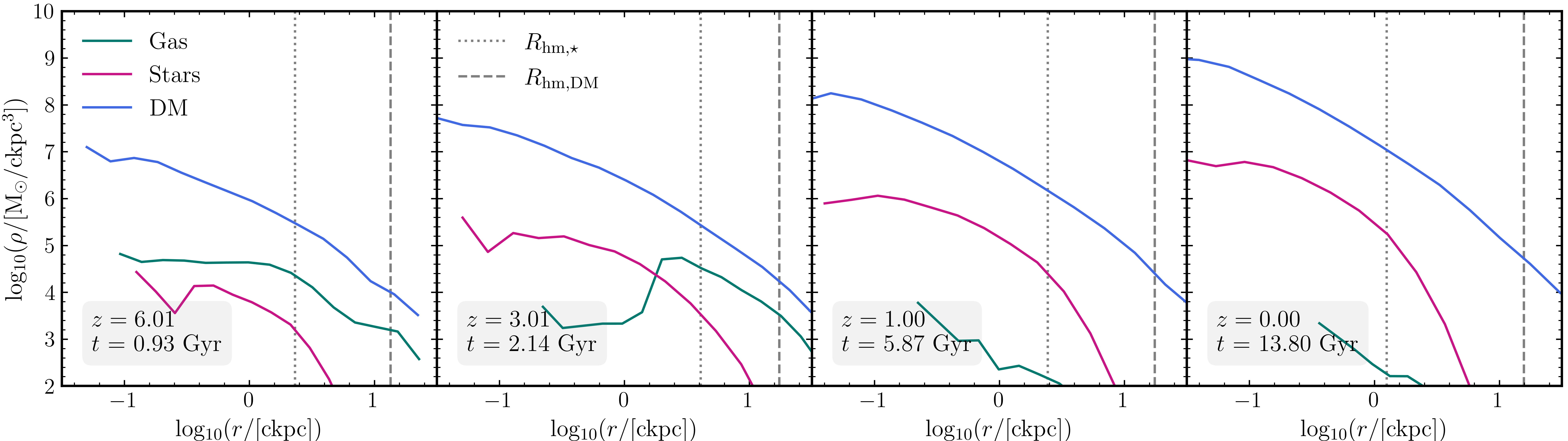} 
    \caption{Radial profiles of the mass density $\rho$ within the virial radius $R_\mathrm{vir}$ for gas (green solid lines), stars (pink solid lines) and DM (blue solid lines), for the \textit{Sink+Low-z} simulation run. To highlight overall trends, we select four representative redshifts, namely $z \sim$ 6, 3, 1 and 0. The 3D stellar half-mass radius $R_\mathrm{hm,\star}$ and the 3D DM half-mass radius $R_\mathrm{hm,DM}$ are indicated as grey dotted and grey dashed lines, respectively. The simulated dwarf is heavily dominated by the DM component at all radii and across cosmic time, which has a much more extended profile compared to its stellar counterpart. In the \textit{Sink+Low-z} run, the gas density distribution is significantly affected by efficient AGN feedback, which expels gas from the central regions of the dwarf, resulting in the observed depletion from $z\sim1$ onwards.} 
    \label{fig: 3D profiles DM vs stars vs gas}
\end{figure*}

To visualise the contribution of the different galaxy components to the mass budget across cosmic time, in Figure~\ref{fig: 3D profiles DM vs stars vs gas} we construct radial profiles of the 3D mass density $\rho$ within the virial radius $R_\mathrm{vir}$ for gas (green solid lines), stars (pink solid lines) and DM (blue solid lines) for the \textit{Sink+Low-z} simulation run. The profiles are plotted at four representative redshifts, namely $z\sim$ 6,  3, 1 and 0, in order to illustrate the key phases of their evolution across cosmic time. For each component, the profiles are constructed as follows: we first find all gas cells or particles within a sphere defined by $R_\mathrm{vir}$ and centred on the particle (or cell) with the minimum gravitational potential energy, at the time the halo is considered. Then, we select fifteen equally-spaced logarithmic radial shells between $r=0$ and $r=R_\mathrm{vir}$. For each shell, the density $\rho$ is computed by dividing the total mass within the shell by the volume of the shell. Note that we do not plot data points corresponding to empty shells. Additionally, at each redshift we plot the 3D stellar half-mass radius $R_\mathrm{hm,\star}$ and the 3D DM half-mass radius $R_\mathrm{hm,DM}$ as grey dotted and grey dashed lines, respectively. 

From this analysis, it emerges that the DM component dominates over the stellar and gaseous ones at all radii and across cosmic time, exhibiting a considerably more extended distribution compared to its stellar counterpart, whose half-mass radius is consistently smaller than that of DM. Moreover, the gas profiles show the strongest evolution with redshift, as efficient AGN feedback depletes the central gas reservoir at $z \gtrsim 1$ (see Figure~\ref{fig: cosmic evolution}). 

Note that these results do not change appreciably across simulation runs, with the exception of the gas profiles, which are naturally sensitive to the specific BH accretion prescription. 


\subsection{Structural properties of the dwarf galaxy for the \textit{Sink+Low-z} run}
\label{sec: violin plots appendix}

To provide a more detailed view of the structural properties of our simulated dwarf in the \textit{Sink+Low-z} run, Figure~\ref{fig: q, n, R_e violins} displays the full distribution of the effective radius $R_\mathrm{e}$, Sérsic index $n$, and axis ratio $q$ at four representative redshifts ($z \sim 6, 3, 1, 0$) for the \textit{Sink+Low-z} run. These distributions are constructed from 100 random lines-of-sight for each snapshot. As discussed in Section~\ref{sec: Mdyn estimators}, the stellar distributions remain relatively compact at high redshift and gradually increase in size toward lower redshift, with the median value of $R_\mathrm{e}$ ranging from $\sim 0.2$~kpc at $z \sim 6$ to $\sim 1$~kpc at $z \sim 0$. The Sérsic index $n$ shows a modest increase over time, indicating that the stellar profiles become slightly more concentrated at later times. The axis ratios remain high ($q \gtrsim 0.7$) at all redshifts, confirming that the dwarfs maintain nearly spherical stellar distributions throughout their evolution. The shaded regions in each panel denote the full spread of values, while the horizontal lines mark the median and the 16th-84th percentile range.

\begin{figure}
    \includegraphics[width=\linewidth]{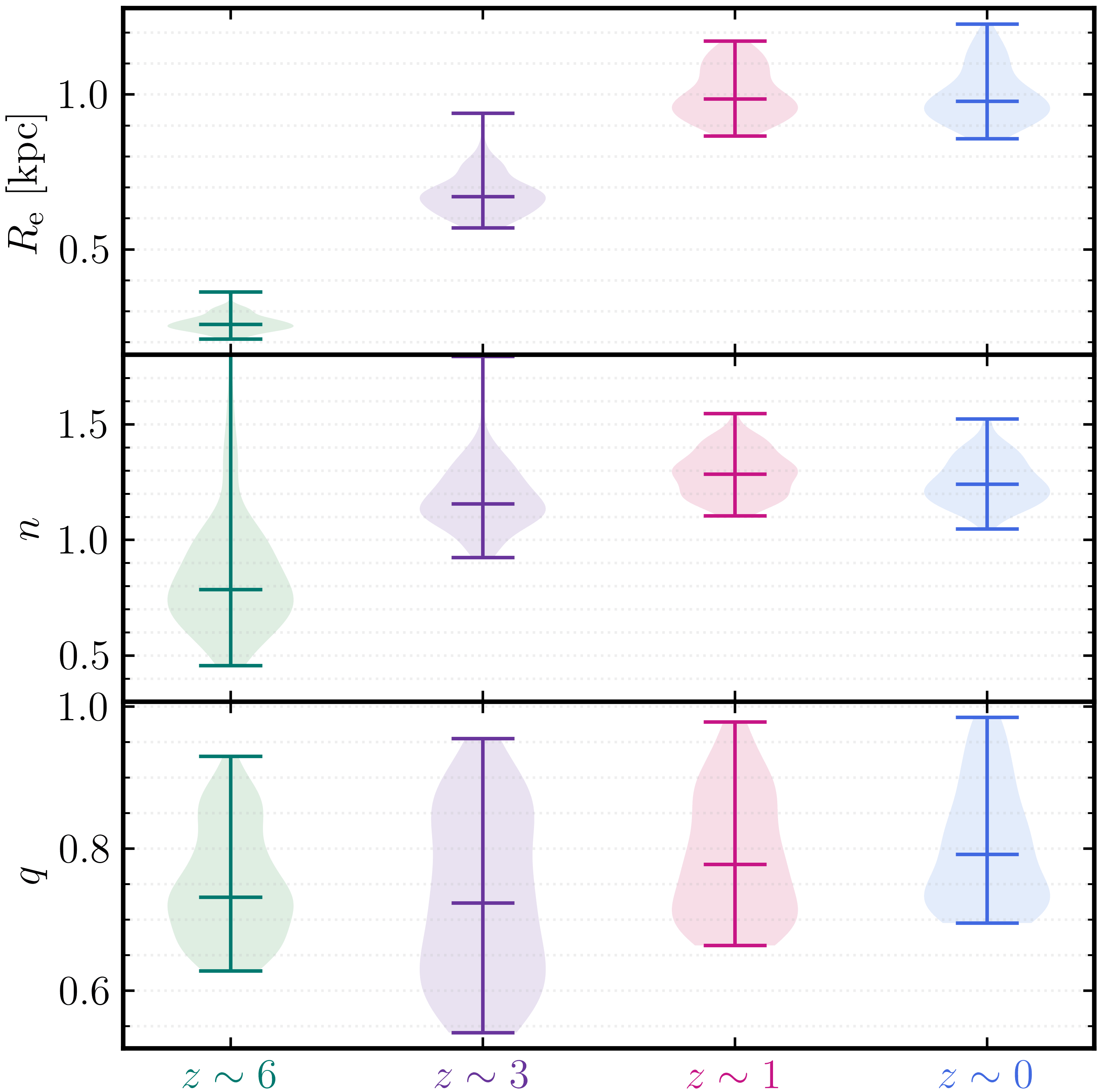} 
        \caption{Distributions of the effective radius $R_\mathrm{e}$, Sérsic index $n$ and axis ratio $q$ for four representative redshifts, i.e. $z \sim 6, 3, 1$ and $0$ for the \textit{Sink+Low-z} run. }
    \label{fig: q, n, R_e violins}
\end{figure}


\subsection{Cosmic evolution of the dynamical mass for all simulation runs}
\label{sec: virial estimator validation appendix}

\begin{figure*}
    \centering
    \includegraphics[width=\textwidth]{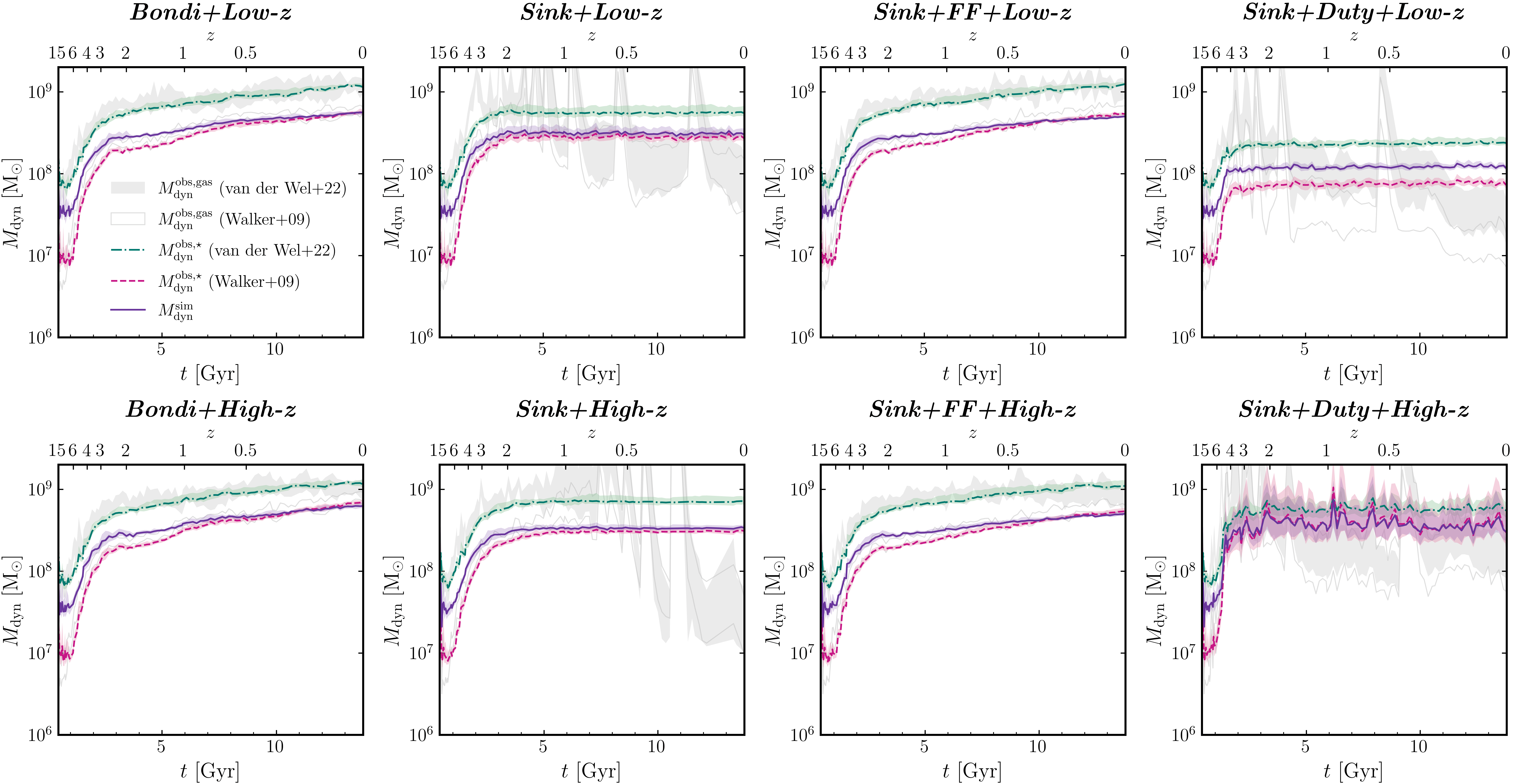} 
        \caption{
        Evolution of the dynamical mass $M_\mathrm{dyn}$ as a function of cosmic time for all simulation runs. The dynamical mass is defined as $M_\mathrm{dyn} \equiv 2 \times M(<R_\mathrm{e})$, where $R_\mathrm{e}$ is the 2D projected stellar half-mass radius. For each simulation and snapshot, we follow the procedure described in Section~\ref{sec: Mdyn estimators}. We compare the true dynamical mass measured directly from the simulations, $M^\mathrm{sim}_\mathrm{dyn}$ (purple solid line), with dynamical masses inferred using the virial estimators of \citet{walker2009universal} and \citet{van2022mass}. When setting $\sigma \equiv \sigma_\star$ in Equation~(\ref{eq: Walker M_dyn}) and Equation~(\ref{eq: vdW M_dyn}), the inferred masses are denoted as $M^\mathrm{obs,\star}_\mathrm{dyn}$ (pink dashed and green dash-dotted lines, respectively). For comparison, we also show results obtained by setting $\sigma \equiv \sigma_\mathrm{gas}$, denoted as $M^\mathrm{obs,gas}_\mathrm{dyn}$ (shown as filled light-grey bands for the \citet{van2022mass} estimator and as unfilled bands bounded by light-grey lines for the \citet{walker2009universal} estimator). Shaded regions indicate the 16th--84th percentile range across the 100 random projections at each snapshot.
    }
    \label{fig: M_dyn vs t all sims}
\end{figure*}

In this section, we repeat the analysis described in Section~\ref{sec: Mdyn estimators} for all simulation runs. Figure~\ref{fig: M_dyn vs t all sims} shows the evolution of the dynamical mass, $M_\mathrm{dyn} \equiv 2 \times M(<R_\mathrm{e})$, where $R_\mathrm{e}$ is the 2D projected stellar half-mass radius obtained from Sérsic fits to the stellar surface mass density profile across 100 random projections at each snapshot.
We compute $M_\mathrm{dyn}$ in three ways. The median true dynamical mass measured directly from the simulations, $M^\mathrm{sim}_\mathrm{dyn}$, is shown as a solid purple line. We then infer dynamical masses using the virial estimators of \citet{walker2009universal} and \citet{van2022mass}. When setting $\sigma \equiv \sigma_\star$ in Equation~(\ref{eq: Walker M_dyn}) and Equation~(\ref{eq: vdW M_dyn}), the corresponding median estimates, $M^\mathrm{obs,\star}_\mathrm{dyn}$, are shown as pink dashed and green dash-dotted lines, respectively, with shaded bands indicating the 16th–84th percentile range across the 100 random projections. For comparison, we also show dynamical masses derived using the gas velocity dispersion, $\sigma \equiv \sigma_\mathrm{gas}$, with the 16th–84th percentile ranges indicated as filled light-grey shaded bands for the \citet{van2022mass} estimator and as unfilled bands bounded by light-grey lines for the \citet{walker2009universal} estimator. 

From this analysis, it first emerges that the systematic overestimation of $M_\mathrm{dyn}$ using the \citet{van2022mass} virial estimator is consistent across all simulation runs, regardless of the specific BH accretion or AGN feedback prescription. This behaviour ultimately reflects the limitations of the estimator when applied to galaxies outside its calibration sample, where structural properties within $R_\mathrm{e}$ may differ significantly from those of the galaxies used to derive it. 

Interestingly, in the \textit{Sink+Duty+Low-z} setup, the \citet{walker2009universal} estimator consistently underpredicts the true dynamical mass across cosmic time. This behaviour likely stems from the following reason. Star formation is quenched at $z \gtrsim 3$ (see right-hand panels in Figure~\ref{fig: cosmic evolution}), with subsequent AGN feedback bursts maintaining the quenched state. As a result, the stellar distribution remains largely unchanged from the time of quenching, placing the galaxy in the regime where the \citet{walker2009universal} estimator underestimates the true dynamical mass in the other setups. Crucially, in those other runs the stellar structure evolves over time, allowing the estimator to provide more accurate dynamical mass predictions. 

Moreover, the \textit{Sink+Duty+High-z} run displays a fundamentally different behaviour compared to all other runs. This can be explained by considering the characteristic timing of strong AGN-driven outflows, which are launched already at $z \gtrsim 5$, when the stellar distribution is still in its early assembly stages. In this high-redshift regime, the gaseous component dominates the mass budget over the stellar component within twice the 3D stellar half-mass radius by a factor of $\sim 100$ (see right-hand panels in Figure~\ref{fig: cosmic evolution}). Consequently, as the AGN impulsively expels this gas, it triggers rapid fluctuations in the local gravitational potential, leading to a reconfiguration of the stellar distribution as it adjusts to a shallower gravitational potential. This results in higher median values of $R_\mathrm{e}$ compared to other runs, ultimately yielding higher typical values of the dynamical mass. This structural difference likely also explains the substantially larger scatter in $M_\mathrm{dyn}$ estimates across different projections at a fixed snapshot. Because the system is dynamically reconfigured during its early assembly, projected quantities such as $R_\mathrm{e}$ and velocity dispersion become more sensitive to LOS variations, thereby increasing the dispersion in the inferred dynamical masses.

Finally, in agreement with the findings discussed in Section~\ref{sec: Mdyn estimators}, runs with efficient AGN feedback show that the gaseous component becomes kinematically decoupled from the stellar one as soon as strong AGN-driven outflows are able to develop, making dynamical mass estimates using the gas velocity dispersion highly unreliable in these systems. However, as we do not resolve the multi-phase ISM, we cannot distinguish between the hot, outflowing gas phase and the colder ISM component. This represents a caveat of our analysis, but an accurate modelling of this distinction is beyond the scope of this work. 


\section{Further diagnostics on metal-enriched outflows}
\label{sec: appendix B}

\begin{figure*}
    \centering
    \includegraphics[width=\textwidth]{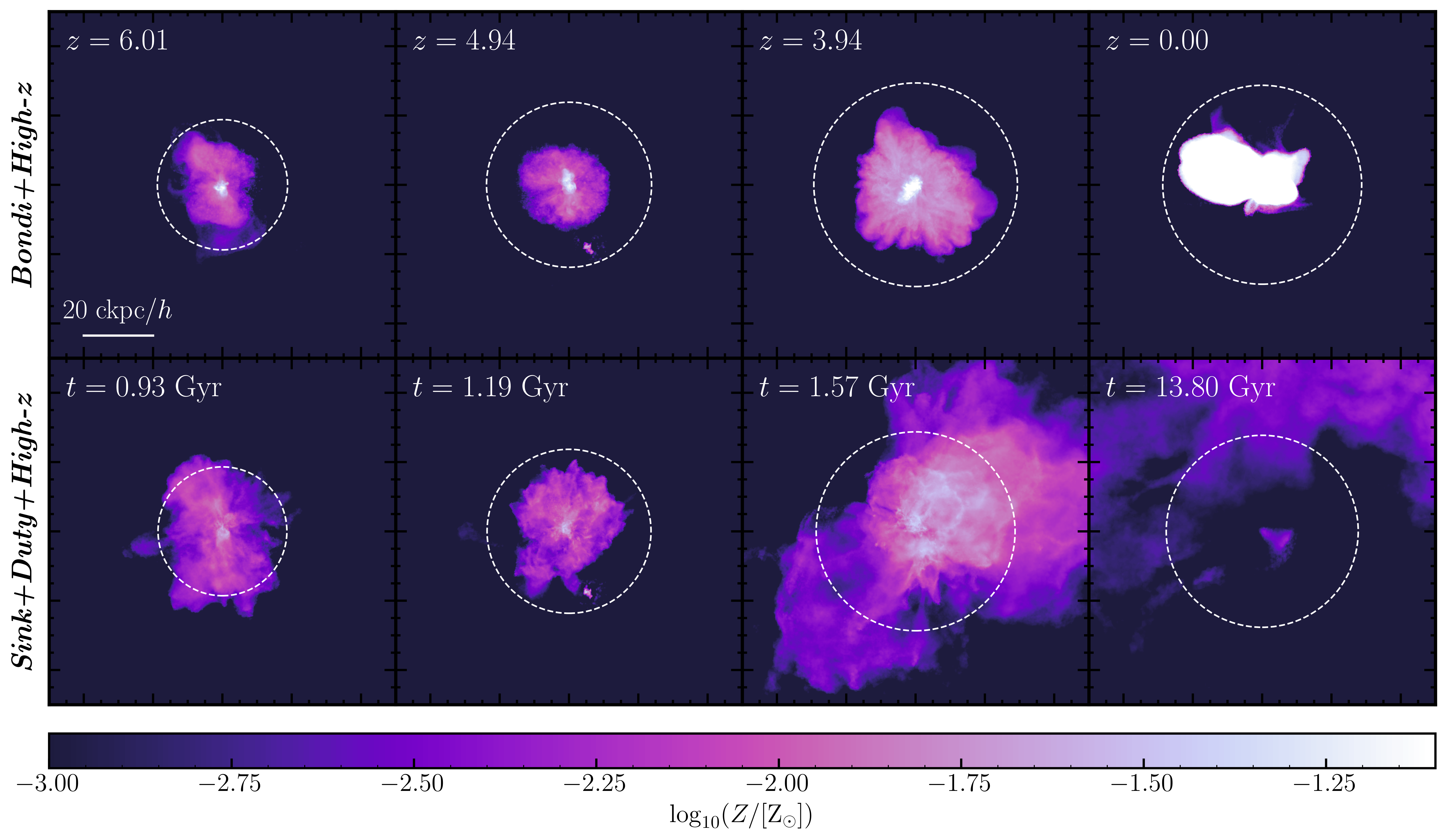} 
    \caption{Density-weighted 2D projections of the gas metallicity for the \textit{Bondi+High-z} (upper panels) and \textit{Sink+Duty+High-z} (lower panels) runs at four representative redshifts, i.e. $z \sim$ 6, 5, 4 and 0.
    The projection dimensions are fixed at $(100 \; \mathrm{ckpc}/h)^3$. The region delimited by the virial radius is indicated by a white dashed circle. The colourbar indicates gas metallicity relative to the solar metallicity. In the \textit{Bondi+High-z} run, metal enrichment remains mostly localised within the innermost regions of the dwarf, as a direct consequence of ongoing star formation activity down to $z=0$. In contrast, the \textit{Sink+Duty+High-z} run displays powerful AGN outflows that deposit metal-enriched gas around and beyond the virial radius at early times ($z \gtrsim 4$). Ultimately, the metal enrichment of the outer CGM serves as a fossil record of past BH accretion and AGN feedback activity, persisting even in systems where the BH is currently dormant and star formation is quenched.}
    \label{fig: high-z runs metals}
\end{figure*}

In this section, we briefly focus on the analysis of two representative runs, namely the \textit{Bondi+High-z} and the \textit{Sink+Duty+High-z} runs, to assess whether metal enrichment in the outer CGM, i.e. around and beyond the virial radius, can serve as a potential tracer of efficient early BH growth and associated AGN feedback. To this end, Figure~\ref{fig: high-z runs metals} shows density-weighted projections of the gas metallicity $Z$ at four representative redshifts, i.e. $z \sim$ 6, 5, 4 and 0, for the \textit{Bondi+High-z} (upper panels) and the \textit{Sink+Duty+High-z} (lower panels) runs. The projection dimensions are fixed at $(100 \; \mathrm{ckpc}/h)^3$ for all panels, and the region delimited by the virial radius is marked as a white dashed circle. 

From this analysis, it emerges that the strong AGN feedback parametrisation within the \textit{Sink+Duty+High-z} setup drives powerful, metal-enriched outflows from the inner, star-forming regions of the dwarf, which begin to propagate beyond the virial radius of the halo at early times ($z 
\sim 6$). Owing to the bursty nature of AGN feedback introduced by the duty cycle, the AGN produces powerful outflows during phases of very efficient BH growth at high redshift ($z \gtrsim 4$; see Figure~\ref{fig: cosmic evolution}), leading to efficient quenching of star formation and halting of further BH accretion. The metals expelled during these episodes remain distributed throughout the outer CGM at later times ($z=0$), even when both star formation and BH growth have effectively ceased. 

In contrast, the \textit{Bondi+High-z} setup displays significantly weaker metal enrichment beyond the virial radius. Although some SN-driven outflows are present in the vicinity of the BH, they are notably slower and less energetic compared to the \textit{Sink+Duty+High-z} setup. As a result, the AGN is unable to transport metal-enriched gas to the outer CGM, highlighting the sensitivity of large-scale metal enrichment to the efficiency and timing of AGN feedback. 

Overall, these findings suggest that metal enrichment around and beyond the virial radius can serve as a long-lived signature of early, efficient BH growth and AGN feedback activity. In particular, compared to the \textit{Sink+Low-z} setup analysed in Section~\ref{sec: outflows}, the rapid, early growth of the BH in the \textit{Sink+Duty+High-z} run drives powerful outflows that deposit metals in the outer CGM at earlier times ($z \gtrsim 4$), showing that the epoch of metal enrichment depends sensitively on the timing of AGN activity. Significant metal enrichment in the outer CGM may therefore encode a fossil record of the past BH growth and AGN activity, even in systems where the BH is currently dormant and star formation is quenched. Nevertheless, due to the extremely low column densities of the outflowing gas in our simulated system, this material is unlikely to be observable with current facilities.

\bsp	
\label{lastpage}
\end{document}